\documentclass[twocolumn]{aastex631}
\usepackage{amsmath}
\usepackage{subfigure}

\newcommand{\DCp}{${\Delta \hat C_{P,{\rm rxn}}}\ $}
\newcommand{\DCpR}{${\Delta \hat C_{P,{\rm rxn}}}/R\ $}
\newcommand{\DV}{$\Delta \hat V_{\rm cond, rxn}\ $}

\newcommand{\DG}{$\Delta \hat G^\circ _{{\rm{rxn}}}\ $}

\shorttitle{Atmosphere-Mantle-Core Connection}
\shortauthors{Schlichting \& Young}

\newcounter{rxn}
\newenvironment{rxn}{%
\refstepcounter{rxn}\equation}
{ \tag{R\arabic {rxn}}\endequation}

\begin{document}

\title{Chemical equilibrium between Cores, Mantles, and Atmospheres of Super-Earths and Sub-Neptunes, and Implications for their Compositions, Interiors, and Evolution}

\correspondingauthor{Hilke E. Schlichting}
\email{hilke@ucla.edu}

\correspondingauthor{Edward D. Young}
\email{eyoung@epss.ucla.edu}

\author[0000-0002-0786-7307]{Hilke E. Schlichting}
\affiliation{Department of Earth, Planetary, and Space Sciences\\
University of California, Los Angeles\\
Los Angeles, CA 90095, USA}

\author{Edward D. Young}
\affiliation{Department of Earth, Planetary, and Space Sciences\\
University of California, Los Angeles\\
Los Angeles, CA 90095, USA}

\begin{abstract}
\begin{ed}
We investigate equilibrium chemistry between molten metal and silicate, and a hydrogen-rich envelope using 18 independent reactions among 25  phase components for sub-Neptune-like exoplanets. Both reactive metal and unreactive metal sequestered in an isolated core are modeled. The overarching effects of equilibration are oxidation of the envelope and reduction of the mantle and core. Hydrogen and oxygen typically comprise significant fractions of metal cores at chemical equilibrium, leading to density deficits that offer a possible alternative explanation for the low densities of the Trappist-1 planets. Reactions with the magma ocean produce significant amounts of SiO and H$_2$O in the envelopes directly above the magma ocean. Molar concentrations in the envelopes of planets with reactive metal are H$_2>\rm{SiO}>CO\sim\rm{Na}\sim\rm{Mg}>\rm{H}_2\rm{O}>>\rm{CO}_2\sim\rm{CH}_4>>\rm{O}_2$ while for the unreactive metal case H$_2$O becomes the second most abundant species, after H$_2$, providing an arbiter for the two scenarios amenable to observation. The water abundances in the atmospheres exceed those in the mantles by at least an order of magnitude in both scenarios.  The water concentrations in the silicate mantles are $\sim 0.01$ wt\% and $\sim 0.1$ wt\% in the reactive and unreactive metal core cases, respectively, limiting the H$_2$O that might be outgassed in a future super-Earth. Less dissolved water in the reactive core case is due to sequestration of H and O in the Fe-rich metal.  The total hydrogen budget of most sub-Neptunes can be, to first order, estimated from their atmospheres alone, as the atmospheres typically contain more than 90\% of all H.
\end{ed}
\end{abstract}



\section{Introduction} \label{sec:intro}

Super-Earths and sub-Neptunes are the most abundant exoplanets discovered to date \citep[e.g.][]{fressin2013a}. Super-Earths constitute a population of smaller, rocky exoplanets, consistent with roughly Earth-like compositions,  while sub-Neptunes constitute a population of larger exoplanets with significant H/He envelopes that contain a few percent of their total mass. 

Recent models of atmospheric evolution and erosion by core-powered mass-loss \citep[e.g.][]{gupta2019a,gupta2020a} and/or photo-evaporation \citep[e.g.][]{owen2017a,rogersj2020a}  suggest that these two populations of exoplanets were born as one. In these models, close-in, lower mass planets lose their hydrogen-helium envelopes and become super-Earths whereas more-massive, longer-period planets retain primordial envelopes and remain sub-Neptunes, thereby explaining the radius valley observed in the exoplanet size distribution \citep[e.g.][]{fulton2017a,eylen2018a,martinez2019a}.

The term ``core" is used in exoplanet research usually to refer to the condensed, non-atmosphere portion of the planet (e.g. core-mass distribution, core-powered mass-loss etc.). In planetary science, ``core" usually refers to the metallic central mass of a differentiated planet. In this work, we follow the planetary science definition of ``core" and refer to the non-atmosphere portion of the planet as either ``core and mantle" or "condensed part of the planet".  \begin{ed} We use the terms ``atmosphere" and ``envelope" interchangeably, recognizing that under the conditions explored here, the phase  boundary between the supercritical H-rich envelope and the magma ocean may not be sharp.\end{ed}

The accretion of solids by a planet forming in a gas disk will be accompanied by the accretion of a hydrogen-dominated envelope once a planetary embryo has reached a mass sufficiently large such that its Bondi radius exceeds the radius of its rocky body. 
As the first layers of gas are accreted, they cool and contract over time, making room for additional gas to be added to the outer regions of the Bondi sphere. Therefore, over typical disk lifetimes, a planet of a few Earth masses embedded in a gas disk can accrete several percent of its total mass in gas \citep[e.g.][]{lee2015a,ginzburg2016a}. The assembly of rocky bodies results in large temperatures as gravitational binding energy is converted into heat. As a result, the
temperature of sub-Neptune interiors is likely dictated by the maximum temperature that still allows for the accretion of H/He envelopes, which corresponds to about 10,000K and more for planets with masses between Earth and Neptune \citep{ginzburg2016a}. The thermal insulation provided by the dense, optically thick, primary atmosphere is sufficient to slow cooling and solidification of the magma oceans over Gyr timescales \citep[e.g.][]{lopez2014a,ginzburg2016a}.  The boundary of the envelope and condensed interior of many sub-Neptunes is therefore likely composed of molten rock immediately beneath the overlying atmosphere.  Analogies to this, although imperfect, exist even in our solar system. The rocky interiors of the misnamed “ice giants” of our solar system, Neptune and Uranus, have temperatures approaching about 7000K, which are masked by radiative temperatures at the top of their atmospheres of only about 60K \citep{Helled2020b}. 

In this work, we aim to understand how these primary, hydrogen-rich atmospheres influence the physical and chemical evolution of super-Earths and sub-Neptunes.  
Since for most exoplanets we will only be able to probe the chemical compositions of their atmospheres, understanding their core-mantle-atmosphere interactions will be crucial for correctly inferring the compositions and chemistries of their underlying mantles and cores, which make up the bulk of these planets by mass. The interface between the surface of the hot rocky body and the overlying dense atmosphere is increasingly recognized as one of the keys to understand the composition, chemical properties and evolution of super-Earths and sub-Neptunes \citep[e.g.][]{ginzburg2016a,Chachan2018,Kite2019,Kite_2020b,Lichtenberg2021a}.

Past works exploring this problem typically only focused on one, two or three chemical reactions \citep[e.g.][]{Chachan2018,Kite2019,Lichtenberg2021b}. Here we use a basis set of 18 reactions and 25 phase components depicting chemical equilibrium between the atmospheres, silicate mantles, and  iron-rich metal cores. These reactions provide a means of characterizing the links between the chemical evolution of  primary atmosphere and the underlying mantle and core. We focus on the distribution of H and omit for the time being He from our analysis due to its inert nature with respect to chemical reactions and much reduced abundance compared to H. 

This paper is structured as follows: In section 2, we present our chemical reaction space and introduce the system of equations that we solve. In section 3 we present the salient features of our results. Section 4 explores the consequences of our results for observations of super-Earths and sub-Neptunes and their evolution.  We summarize our findings in section 5.

\section{Equilibrium reaction space}

\subsection{General approach}

Our goal is to create an equilibrium thermodynamic model for a sub-Neptune-like exoplanet to account for  chemical interactions between a metal core, a silicate dominated mantle and a hydrogen-rich atmosphere. To account for the possibility of differentiation and limited bulk transport in the planet \citep[e.g.][]{Lichtenberg2021b}, we will be investigating, as end-member scenarios, both a ``reactive metal" case, in which the metal, perhaps comprising a differentiated core, participates fully in the equilibrium chemistry, and an ``unreactive metal" case, where metal is sequestered into a core that is isolated from interactions with the mantle and atmosphere. Unlike kinetic reaction networks in which the rates of every relevant chemical pathway should be specified, full  descriptions of systems in chemical equilibrium require only a basis set of reactions; equilibrium for all other relevant reactions are defined by linear combinations of those comprising the basis set. Our goal is therefore met by constructing a suitable set of reaction basis vectors that span the reaction space. 

The planets we consider are composed of a hydrogen dominated atmosphere surrounding a magma ocean mantle and metal core.  We constructed a basis set of reactions that is intended to be simple enough to interpret, yet sufficient to capture the fundamental aspects of chemical interactions between the atmosphere, mantle, and  metal core (if there is one). The condition for chemical equilibrium is a zero sum of the product of the chemical potentials of species $i$, $\mu_i$, in a chemical reaction and their stoichiometric coefficients in the reaction, $\nu_i$. The conditions for chemical equilibrium are therefore $\sum_i \nu_i \mu_i = 0$ for each reaction comprising the basis set. Chemical potentials are defined specific to a phase (silicate, metal, or atmosphere in our case).  We specify as an initial condition that the planets are composed of $N_{\rm metal}$, $N_{\rm silicate}$, and $N_{\rm atm}$ moles of metal, silicate, and atmosphere, respectively, but the proportions of each phase change in response to reaching chemical equilibrium.  The chemical species constituting each of these phases are referred to as ``phase components''.  The minimum number of chemical constituents necessary to describe the compositions of the phase components are the ``system components.'' 
There are more phase components than system components.  For phase components $n_{\phi}$ in number, and $n_{\rm s}$ number of system components, there are $n_{\phi} - n_{\rm s} = n_{\rm r}$ linearly independent reactions that describe the chemical interactions among the components.  These $n_{\rm r}$ reactions comprise the basis set necessary to describe thermodynamic equilibrium among the phase components.  The system of equations to be solved to define the equilibrium state consists of the conditions for equilibrium for the $n_{\rm r}$ reactions, a set of $n_s$ mass balance equations specifying the constant mass, or moles, of each of the system components of the planet, and summing constraints for the mole fractions of each phase.  

\subsection{Constructing the system of equations}
We derive the basis set of reactions in three parts: 1) reactions among constituents of both the metal core and silicate mantle; 2) reactions among constituents of the atmosphere; and 3) exchange between the atmosphere, mantle and core.

The 13 phase components making up the liquid mantle and core include:  MgO, SiO$_2$, MgSiO$_3$, FeO, FeSiO$_3$, Na$_2$O, Na$_2$SiO$_3$, Fe$_{\rm metal}$, Si$_{\rm metal}$, O$_{\rm metal}$, H$_{\rm metal}$, H$_{2{\rm silicate}}$, and H$_2$O$_{\rm silicate}$. Why these components?  They were chosen to allow evaporation from the magma ocean to be written in term of reactions involving the simple oxides (MgO, SiO$_2$, FeO, Na$_2$O) but allowing for the fact that the thermodynamic activities of these components are known to be quite low compared with their concentrations in the liquid.  Recalling that activity of a component $i$ is the effective concentration given by $a_i = \gamma_i x_i$, where $x_i$ is mole fraction (concentration) of the component in the melt and $\gamma_i$ is the activity coefficient (a factor that accommodates the difference between $a_i$ and $x_i$), there are two ways to account for this low activity relative to mole fractions.  One is to assign a low value to the activity coefficient $\gamma_i$, but these are not well known. The other is to consider that the way that Mg, for example, exists in a silicate liquid is not mainly as MgO, but rather as MgSiO$_3$ and other similar silicate mineral-like species through reactions like MgO + SiO$_2$ = MgSiO$_3$ in the melt.  The activities of the simple oxides can be accounted for, to first order, by such reactions.  Here we include at least one other liquid component for Mg, Fe,  Si, and Na to account for this phenomenon.  This has the added advantage that we can treat the mixing in the liquid as ideal, meaning $a_i = x_i$.  
We include the metallic components to allow for the importance of reactions between silicate melt and metal melt for controlling the oxidation states of both.  Transfer of O, Si, and H in and out of Fe-rich metal is an important aspect of the interaction between the atmosphere and the core when the atmosphere is rich in hydrogen.  This is true regardless of whether the metal is entrained in a vigorously convecting mantle or coalesced to the center of the planet \citep[e.g.][]{Lichtenberg2021b}. Finally, we include H$_2$ and H$_2$O dissolved in the silicate liquid because hydrogen and water will play a critical role in mediating oxidation and reduction reactions in the core. Some fraction of water speciates to OH in the melt, but this is ignored for now and is unlikely to be critical to the results reported here. Similarly, there is a rich chemistry to be explored in the atmospheres of exoplanets, including reaction networks driven by photochemistry, but the components considered here should capture the most fundamental features of core-mantle-atmosphere exchange.

The 13 phase components of the molten core and mantle are described by 6 system components: Si, Mg, O, Fe, H, and Na.  We derive the linearly independent set of reactions that can occur in the liquid core by writing the equations for each phase component in terms of the system components, as in 1MgO = 1Mg + 1O, or 1MgSiO$_3$ = 1Mg + 1Si + 3O, and so forth.  By doing so one arrives at a set of equations describing the compositions of each phase component, which in matrix form is

\begin{equation}
\left[ \begin{array}{l}
{\rm{MgO}}\\
{\rm{Si}}{{\rm{O}}_{\rm{2}}}\\
{\rm{MgSi}}{{\rm{O}}_{\rm{3}}}\\
{\rm{FeO}}\\
{\rm{FeSi}}{{\rm{O}}_{\rm{3}}}\\
{\rm{N}}{{\rm{a}}_{\rm{2}}}{\rm{O}}\\
{\rm{N}}{{\rm{a}}_{\rm{2}}}{\rm{Si}}{{\rm{O}}_{\rm{3}}}\\
{\rm{F}}{{\rm{e}}_{{\rm{metal}}}}\\
{\rm{S}}{{\rm{i}}_{{\rm{metal}}}}\\
{{\rm{O}}_{{\rm{metal}}}}\\
{{\rm{H}}_{{\rm{metal}}}}\\
{{\rm{H}}_{{\rm{2,}}}}_{{\rm{silicate}}}\\
{{\rm{H}}_{\rm{2}}}{{\rm{O}}_{{\rm{silicate}}}}
\end{array} \right]\quad  = \quad 
\left[ {\begin{array}{*{20}{c}}
0&1&1&0&0&0\\
1&0&2&0&0&0\\
1&1&3&0&0&0\\
0&0&1&1&0&0\\
1&0&3&1&0&0\\
0&0&1&0&0&2\\
1&0&3&0&0&2\\
0&0&0&1&0&0\\
1&0&0&0&0&0\\
0&0&1&0&0&0\\
0&0&0&0&1&0\\
0&0&0&0&2&0\\
0&0&1&0&2&0
\end{array}} \right]\;\quad 
\left[ \begin{array}{l}
{\rm{Si}}\\
{\rm{Mg}}\\
{\rm{O}}\\
{\rm{Fe}}\\
{\rm{H}}\\
{\rm{Na}}\\
\end{array} \right].\
\end{equation}

 \smallskip

\noindent Row reduction of the coefficient matrix exposes the linear dependencies (zero rows) among the phase components MgO, $\rm{SiO_2}$, and so forth.  These dependencies are the basis-set reactions for the mantle and core,  in this case $13-6 = 7$ in number.  In this case there are therefore $7$ linearly independent reactions among the 13 phase components of the mantle and core.  These reactions, all occurring in the liquid silicate or between the immiscible liquid silicate and liquid metal, include:  speciation of Na$_2$O in melt

\begin{rxn}
{\rm{N}}{{\rm{a}}_{\rm{2}}}{\rm{Si}}{{\rm{O}}_{\rm{3}}} \mathbin{\lower.3ex\hbox{$\buildrel\textstyle\rightarrow\over
{\smash{\leftarrow}\vphantom{_{\vbox to.5ex{\vss}}}}$}} {\rm{N}}{{\rm{a}}_{\rm{2}}}{\rm{O + Si}}{{\rm{O}}_{\rm{2}}};
\end{rxn}

\noindent an oxidation and reduction exchange reaction between Fe in metal and Si in metal mediated by the silicate melt, 

\begin{rxn}
{\rm{1/2}}\;{\rm{Si}}{{\rm{O}}_{\rm{2}}}{\rm{ +  F}}{{\rm{e}}_{{\rm{metal}}}} \mathbin{\lower.3ex\hbox{$\buildrel\textstyle\rightarrow\over
{\smash{\leftarrow}\vphantom{_{\vbox to.5ex{\vss}}}}$}} {\rm{FeO}}\;{\rm{ + }}\;{\rm{1/2}}\;{\rm{S}}{{\rm{i}}_{{\rm{metal}}}} ;
\end{rxn}

\noindent speciation of Mg in the melt,
\begin{rxn}
{\rm{MgSi}}{{\rm{O}}_{\rm{3}}} \mathbin{\lower.3ex\hbox{$\buildrel\textstyle\rightarrow\over
{\smash{\leftarrow}\vphantom{_{\vbox to.5ex{\vss}}}}$}} {\rm{MgO  +  Si}}{{\rm{O}}_{\rm{2}}} ;
\end{rxn}

\noindent reduction of O in metal balanced by oxidation of Si in metal,
\begin{rxn}
{{\rm{O}}_{{\rm{metal}}}} + {\rm{ 1/2 S}}{{\rm{i}}_{{\rm{metal}}}} \mathbin{\lower.3ex\hbox{$\buildrel\textstyle\rightarrow\over
{\smash{\leftarrow}\vphantom{_{\vbox to.5ex{\vss}}}}$}} {\rm{ 1/2 Si}}{{\rm{O}}_2} ;
\end{rxn}

\noindent exchange of hydrogen between metal and silicate,
\begin{rxn}
{\rm{2 }}{{\rm{H}}_{{\rm{metal}}}}{\rm{ }} \mathbin{\lower.3ex\hbox{$\buildrel\textstyle\rightarrow\over
{\smash{\leftarrow}\vphantom{_{\vbox to.5ex{\vss}}}}$}} {\rm{ }}{{\rm{H}}_{\rm{2}}}_{,{\rm{silicate}}} ;
\end{rxn}

\noindent speciation of FeO in the silicate melt, 
\begin{rxn}
{\rm{FeSi}}{{\rm{O}}_{\rm{3}}}{\rm{ }} \mathbin{\lower.3ex\hbox{$\buildrel\textstyle\rightarrow\over
{\smash{\leftarrow}\vphantom{_{\vbox to.5ex{\vss}}}}$}} {\rm{ FeO  +  Si}}{{\rm{O}}_{\rm{2}}} ;
\end{rxn}

\noindent and oxidation of Si in metal by water in the melt,

\begin{rxn}
{\rm{2 }}{{\rm{H}}_{\rm{2}}}{{\rm{O}}_{{\rm{silicate}}}}{\rm{  +  S}}{{\rm{i}}_{{\rm{metal}}}}{\rm{ }} \mathbin{\lower.3ex\hbox{$\buildrel\textstyle\rightarrow\over
{\smash{\leftarrow}\vphantom{_{\vbox to.5ex{\vss}}}}$}} {\rm{Si}}{{\rm{O}}_{\rm{2}}}{\rm{  +  2 }}{{\rm{H}}_{{\rm{2, silicate}}}} . 
\end{rxn}

\medskip
\noindent We emphasize that other reactions that one might wish to write among these phase components are fully accounted for by reactions R1 through R7.  For example, one may envisage the introduction of Si to the metal by the reduction/oxidation (redox) reaction MgSiO$_3$ + $2/3$Fe$_{\rm metal} \rightleftharpoons$ MgO + $2/3$FeSiO$_3$ + $1/3$Si$_{\rm metal}$.  Equilibrium for this reaction is accounted for since it is described by the basis set as R3 $+$ R2 $-2/3$ R6.

Next we consider the reactions that take place in the atmosphere.  Because CO, an  abundant reservoir of carbon in solar-like gases \citep{France2014}, is likely to be an important constituent in the initial primary atmosphere, we add carbon to our list of system components.  For consistency, we follow the same procedure as we did for the magma ocean, although this result is trivial and could be written {\it a priori}, yielding

\begin{equation}
\left[ \begin{array}{l}
{{\rm{H}}_{{\rm{2, gas}}}}\\
{\rm{C}}{{\rm{O}}_{{\rm{gas}}}}\\
{\rm{C}}{{\rm{O}}_{{\rm{2, gas}}}}\\
{\rm{C}}{{\rm{H}}_{{\rm{4, gas}}}}\\
{{\rm{O}}_{\rm{2}}}_{,{\rm{gas}}}\\
{{\rm{H}}_{\rm{2}}}{{\rm{O}}_{{\rm{gas}}}}
\end{array} \right]\quad  = \quad \left[ {\begin{array}{*{20}{c}}
0&0&2\\
1&1&0\\
1&2&0\\
1&0&4\\
0&2&0\\
0&1&2
\end{array}} \right]\quad \left[ \begin{array}{l}
{\rm{C}}\\
{\rm{O}}\\
{\rm{H}}
\end{array} \right].
\end{equation}

\noindent Row reduction of the above, or in this case simple inspection, shows that for these 6 phase components comprising the atmosphere and the three system components that make up the atmosphere, we have three linearly independent reactions that can take place in the atmosphere: 

\begin{rxn}
{\rm{C}}{{\rm{O}}_{{\rm{gas}}}}{\rm{  +  1/2 }}{{\rm{O}}_{{\rm{2, gas}}}} \mathbin{\lower.3ex\hbox{$\buildrel\textstyle\rightarrow\over
{\smash{\leftarrow}\vphantom{_{\vbox to.5ex{\vss}}}}$}} {\rm{ C}}{{\rm{O}}_{{\rm{2, gas}}}}{\rm{ }},
\end{rxn}

\begin{rxn}
{\rm{C}}{{\rm{H}}_{{\rm{4, gas}}}}{\rm{  +  1/2 }}{{\rm{O}}_{{\rm{2, gas}}}}{\rm{ }} \mathbin{\lower.3ex\hbox{$\buildrel\textstyle\rightarrow\over
{\smash{\leftarrow}\vphantom{_{\vbox to.5ex{\vss}}}}$}} {\rm{ 2 }}{{\rm{H}}_{\rm{2}}}{\rm{  +  CO}},
\end{rxn}

\noindent  and

\begin{rxn}
{{\rm{H}}_{{\rm{2, gas}}}}{\rm{  +  1/2 }}{{\rm{O}}_{{\rm{2, gas}}}}{\rm{ }} \mathbin{\lower.3ex\hbox{$\buildrel\textstyle\rightarrow\over
{\smash{\leftarrow}\vphantom{_{\vbox to.5ex{\vss}}}}$}} {\rm{ }}{{\rm{H}}_{\rm{2}}}{{\rm{O}}_{{\rm{gas}}}}.
\end{rxn}

\medskip

Lastly, to these 10 reactions we add an additional 8 reactions that capture the exchange of atmophile molecules with the magma ocean.  Four correspond to equilibrium vapor pressures of moderately volatile lithophile elements and the other four represent the solubilities of atmophile gases in the magma ocean.  The 8 reactions can be written {\it a priori} and are:

\begin{rxn}
{\rm{FeO }} \mathbin{\lower.3ex\hbox{$\buildrel\textstyle\rightarrow\over
{\smash{\leftarrow}\vphantom{_{\vbox to.5ex{\vss}}}}$}} {\rm{F}}{{\rm{e}}_{{\rm{gas}}}}{\rm{  +  1/2 }}{{\rm{O}}_{{\rm{2,gas}}}},
\end{rxn}

\begin{rxn}
{\rm{MgO }} \mathbin{\lower.3ex\hbox{$\buildrel\textstyle\rightarrow\over
{\smash{\leftarrow}\vphantom{_{\vbox to.5ex{\vss}}}}$}} {\rm{ M}}{{\rm{g}}_{{\rm{gas}}}}{\rm{ +  1/2 }}{{\rm{O}}_{{\rm{2,gas}}}},
\end{rxn}

\begin{rxn}
{\rm{Si}}{{\rm{O}}_{\rm{2}}}{\rm{ }} \mathbin{\lower.3ex\hbox{$\buildrel\textstyle\rightarrow\over
{\smash{\leftarrow}\vphantom{_{\vbox to.5ex{\vss}}}}$}} {\rm{ Si}}{{\rm{O}}_{{\rm{gas}}}}{\rm{  +  1/2 }}{{\rm{O}}_{{\rm{2,gas}}}},
\end{rxn}

\begin{rxn}
{\rm{N}}{{\rm{a}}_{\rm{2}}}{\rm{O }} \mathbin{\lower.3ex\hbox{$\buildrel\textstyle\rightarrow\over
{\smash{\leftarrow}\vphantom{_{\vbox to.5ex{\vss}}}}$}} {\rm{ 2 N}}{{\rm{a}}_{{\rm{gas}}}}{\rm{  +  1/2 }}{{\rm{O}}_{{\rm{2,gas}}}},
\end{rxn}

\begin{rxn}
{{\rm{H}}_{{\rm{2,gas}}}}{\rm{ }} \mathbin{\lower.3ex\hbox{$\buildrel\textstyle\rightarrow\over
{\smash{\leftarrow}\vphantom{_{\vbox to.5ex{\vss}}}}$}} {\rm{ }}{{\rm{H}}_{{\rm{2,silicate}}}},
\end{rxn}

\begin{rxn}
{\rm H_2O_ {\rm{gas }}} \mathbin{\lower.3ex\hbox{$\buildrel\textstyle\rightarrow\over
{\smash{\leftarrow}\vphantom{_{\vbox to.5ex{\vss}}}}$}} {\rm H_2O_ {\rm{, silicate}}}
\end{rxn}

\begin{rxn}
{\rm{C}}{{\rm{O}}_{{\rm{gas}}}}{\rm{ }} \mathbin{\lower.3ex\hbox{$\buildrel\textstyle\rightarrow\over
{\smash{\leftarrow}\vphantom{_{\vbox to.5ex{\vss}}}}$}} {\rm{ C}}{{\rm{O}}_{{\rm{silicate}}}},
\end{rxn}

\noindent and

\begin{rxn}
{\rm{C}}{{\rm{O}}_{{\rm{2, gas}}}}{\rm{ }} \mathbin{\lower.3ex\hbox{$\buildrel\textstyle\rightarrow\over
{\smash{\leftarrow}\vphantom{_{\vbox to.5ex{\vss}}}}$}} {\rm{ C}}{{\rm{O}}_{{\rm{2, silicate}}}}.
\end{rxn}

\noindent Reactions (R11) through (R18) add an additional 6 phase components to the 19 we already have, for a total of 25 phase components. 

We note that previous work on reactions between the core and an H$_2$-rich atmosphere are captured by reactions R1 through R18 here. For example, the reaction FeO$_{\rm silicate}$ + H$_{2,{\rm gas}} \rightleftharpoons {\rm Fe}_{\rm metal}$ + H$_2$O$_{\rm gas}$ examined by \cite{Kite2021} is R15$-$R16$-$R2$-1/2$ R7 in our reaction space. 
  
For each of the 18 reactions among 25 phase components described above there is an equation for the condition for equilibrium.  These equations are of the form  

\begin{equation}
\sum\limits_i {{\nu _i}\ln {x_i} + \left[ {\frac{{\Delta \hat G_{{\rm{rxn}}}^ \circ }}{{RT}} + \sum\limits_g {{\nu _g}\ln (P/{P^ \circ })} } \right]}  = 0.
\label{eqn:f}
\end{equation}

\noindent Here we have used $\mu_i = \Delta \hat G_i^{\rm o} + RT \ln(x_i)$ for the chemical potential of species $i$ corrected for composition where  $\Delta \hat G_i^{\rm o}$ is the Gibbs free energy of formation of $i$ at a standard state of the pure species at temperature and 1 bar pressure, and $R$ is the gas constant.  The sum over index $g$ refers to the gas species only in the reaction and index $i$ refers to all species, including the gases.  We  separated the compositional and pressure effects for the gas species, replacing partial pressure (ideal fugacity), for species $g$, by the product $x_g P$ where $P$ is the total gas pressure and $P^{\rm o}$ is the pressure at standard state (chosen as 1 bar here). We discuss below the effects of excluding the  partial molar volumes of melt species in the pressure corrections. 

To these 18 equations in terms of the 25 mole fractions we add an additional 7 equations that account for the summing constraints for each of the 7 elements making up the planet.  The mass balance equations are of the form

\begin{equation}
{n_s} - \sum\limits_k {\sum\limits_i {{\eta _{s,i,k}}{x_{i,k}}{N_k}} }  = 0, 
\end{equation}

\noindent where $n_s$ is the moles of element $s$ in the planet, $\eta_{s,i,k}$  is the number of moles of element $s$ in component $i$ of phase $k$, $x_{i,k}$ is the mole fraction of component $i$ in phase $k$, and $N_k$ is the moles of phase $k$ (metal, silicate, or atmosphere). The moles of the phases are treated as variables along with the mole fractions for the species, resulting in 28 variables in the system of equations. To  the 25 equations described thus far, we add three additional equations among the mole fractions that state explicitly that the sum of the mole fractions for each phase must be unity.  Our new equations include

\begin{equation}
\begin{array}{l}
1 - {x_{{\rm{MgO}}}} - {x_{{\rm{Si}}{{\rm{O}}_{\rm{2}}}}} - {x_{{\rm{MgSi}}{{\rm{O}}_{\rm{3}}}}} - {x_{{\rm{FeO}}}} - {x_{{\rm{FeSi}}{{\rm{O}}_{\rm{3}}}}} - \\
{x_{{\rm{N}}{{\rm{a}}_{\rm{2}}}{\rm{O}}}} - {x_{{\rm{N}}{{\rm{a}}_{\rm{2}}}{\rm{Si}}{{\rm{O}}_{\rm{3}}}}} - x_{{{\rm{H}}_{\rm{2}}}}^{{\rm{silicate}}} - 
x_{{{\rm{H}}_{\rm{2}}}{\rm{O}}}^{{\rm{silicate}}} - \\
x_{{\rm{CO}}}^{{\rm{silicate}}} - x_{{\rm{C}}{{\rm{O}}_{\rm{2}}}}^{{\rm{silicate}}} = 0,
\end{array}
\end{equation}

for the silicate phase, 

\begin{equation}
1 - x_{{\rm{Fe}}}^{{\rm{metal}}} - x_{{\rm{Si}}}^{{\rm{metal}}} - x_{\rm{O}}^{{\rm{metal}}} - x_{\rm{H}}^{{\rm{metal}}} = 0,
\end{equation}

for the metal phase, and

\begin{equation}
\begin{array}{l}
1 - x_{{\rm{CO}}}^{{\rm{gas}}} - x_{{\rm{C}}{{\rm{O}}_{\rm{2}}}}^{{\rm{gas}}} - x_{{{\rm{O}}_{\rm{2}}}}^{{\rm{gas}}} - x_{{\rm{C}}{{\rm{H}}_{\rm{4}}}}^{{\rm{gas}}} - x_{{{\rm{H}}_{\rm{2}}}}^{{\rm{gas}}} - x_{{{\rm{H}}_{\rm{2}}}{\rm{O}}}^{{\rm{gas}}} - x_{{\rm{Fe}}}^{{\rm{gas}}} - \\
x_{{\rm{Mg}}}^{{\rm{gas}}} - x_{{\rm{SiO}}}^{{\rm{gas}}} -x_{{\rm{Na}}}^{{\rm{gas}}} = 0
\end{array}
\end{equation}

\medskip
\noindent for the atmosphere.  

Finally, we add the atmospheric pressure at the surface of the magma ocean, $P_{\rm surface}$, as an additional variable.  The pressure depends on the mean molecular weight of the atmosphere, $\mu$, which in turn depends on the molecular species comprising the atmosphere.  We must therefore solve for the surface pressure simultaneously with the chemical speciation equations.  This is accomplished by adding the equation  

\begin{equation}
\left( \frac{P_{\rm surface}}{1 {\rm bar}} \right) \simeq 1 \times 10^6 \frac{M_{\rm atm}}{M_{\rm p}} \left( \frac{M_{\rm p}}{M_{\Earth}} \right)^{2/3}
\end{equation}

\noindent to the system of equations to be solved, where the mass of the atmosphere, $M_{\rm atm}$, is obtained from the grams per mole ($\mu$) and the moles of atmosphere, $N_{\rm atm}$.

\begin{ed}
The 29 non-linear equations in 29 variables were solved using simulated annealing followed by Markov chain Monte Carlo (MCMC) sampling.  We used the Python implementation of simulated annealing  \citep[dual\_annealing,][]{Xiang_1997} and the  MIT Python implementation of ensemble MCMC \citep[emcee,][]{Foreman2019} based on methods described by \cite{Goodman2010}.   The combination of simulated annealing followed by MCMC sampling avoids becoming trapped in the large number of local minima (100's on a typical search).  Our criteria for acceptable solutions included: 1) equilibrium constants accurate to $< 0.5\%$; 2) mass balance  satisfied to $\le 0.006 \%$  for each element; 3) summing constraints for the mole fractions for each phase being satisfied to the 5th decimal digit or better; and 4) pressure achieving an accuracy of $< 10\%$ with most solutions having an accuracy of $< 1\%$.  
\end{ed}

\subsection{Chemical Thermodynamics at high  P and T}\label{Thermo}
\begin{ed}
We make several simplifying assumptions in this analysis in order to make the problem of characterizing the chemical equilibrium state of an entire planet more tractable.  These include using the temperature at the surface of the magma ocean to evaluate the equilibrium constants, confining pressure corrections to equilibrium constants to the atmophile components only (Equation \ref{eqn:f}), and the use of ideal gas behavior for the hydrogen-rich atmospheres.  Here we assess the likely impact of these simplifications on the accuracy of our results.

Using the equations of state in this paper, the pressure at the boundary between a central metal core and the silicate mantle for the 4 $M_\earth$ planets we consider should be $\sim 450$ GPa. The isoentropic adiabat of \cite{Wilson2021} suggests that the temperature at this location could be as high as 14,000 K with an estimated maximum surface $T$ of 6000 K.  The conditions considered here are well above the liquidus for silicate and metal, so our planet beneath the atmosphere is considered to be entirely molten.  Pressure effects on intra-melt reactions are mitigated by the fact that as pressure increases beyond a few GPa, the thermochemistry of the melt species become less pressure dependent.  Efficient atomic packing causes many thermodymamic parameters to level off above about 100 GPa \citep{Stixrude2005}.  Partial molar volumes of silicate molten species decrease markedly from  0 to 20 GPa, with a more gradual decrease at higher pressures \citep{DeKoker2009}.  Similarly, differences in heat capacities among melt species are small enough that the effects of temperature tend to offset the pressure effects.  We explore these claims with a specific example below. 

The relationship between  equilibrium constants that are the basis for Equation \ref{eqn:f} and the equations of state for the species in the reactions is embodied by the pressure and temperature dependence of the Gibbs free energy of reaction according to

\begin{equation}
\begin{aligned}
-&RT\ln{K_{\rm eq}}=
{\Delta \hat G_{{\rm{rxn}}}^ \circ }+\int_{T_{\rm ref}}^T {\Delta \hat C_{P,{\rm rxn}}}dT \\
&-T\int_{T_{\rm ref}}^T\frac{\Delta \hat C_{P,{\rm rxn}}}{T}dT  +\Delta \hat V_{\rm cond, rxn}(P-P^\circ),
\end{aligned}
\label{eqn:TandP}
\end{equation}
where ${\Delta \hat C_{P,{\rm rxn}}}$ is the difference in molar isobaric heat capacities between products and reactants, $\Delta \hat V_{\rm cond, rxn}$ is the difference in the partial molar volumes of product and reactant condensed species (pressure effects for the gaseous or supercritical fluid species are now embedded in the equilibrium constants through the fugacities in Equation \ref{eqn:TandP}, for convenience), and $T_{\rm ref}$ and $P^\circ$ are the reference temperature and pressure used to calculate \DG (e.g., 3500 K to 6000 K and 1 bar, respectively).  Equation \ref{eqn:TandP} emphasizes that the temperature and pressure dependence of the equilibrium constants rely on small differences between large numbers.  

At temperatures above their Debye temperatures, molar isochoric heat capacities of crystalline solids approach the Dulong-Petit limit where $\hat C_{\rm V} = 3nR$ and $n$ is the number of atoms per formula unit.  For melt species, additional vibrational modes result in heat capacities that exceed this limit by approximately $30\%$, with $\hat C_{\rm V} \sim 4nR$ and $C_{\rm P}/C_{\rm V}$ ratios of between $\sim 1.1$ to $1.3$ \citep{DeKoker2009}.  To the extent that the heat capacities for all melt species approach their high-T limits of $\sim 4nR$ ,  the \DCp for balanced intra-melt speciation reactions should be zero (the $n$'s cancel).  However, in detail, the deviations from the high-$T$ limit of $\sim 4nR$ differ slightly among the various species.  As an example, we examine the effects of these differences for reaction R3.  We use this reaction because the thermodynamic parameters as functions of $T$ and $P$ for the Mg-endmember melt species, unlike many other species of interest, are well characterized. 

 There is little lost by assuming that \DCp values are constant over the ranges of high temperatures considered here.  Equation \ref{eqn:TandP} thus simplifies to
 
 \begin{equation}
\begin{aligned}
-&\ln{K_{\rm eq}}=
\frac{{\Delta \hat G^\circ_{{\rm{rxn}}}}}{RT}+\frac{\Delta \hat C_{P,{\rm rxn}}}{RT}(T-T_{\rm ref}) \\
&-\frac{\Delta \hat C_{P,{\rm rxn}}}{R}\ln(T/T_{\rm ref})  +\frac{\Delta \hat V_{\rm cond, rxn}}{RT}(P-P^\circ).
\end{aligned}
\label{eqn:TandPsimple}
\end{equation}
The effect on the equilibrium constant for reaction R3 can therefore be evaluated using the isobaric heat capacities reported by \cite{DeKoker2009} over a range of $T$ from 1773 K to 3000K.  These are $\hat C_P$(MgSiO$_3$) $= 4.6nR$, $\hat C_P$(MgO) $= 4.7nR$, and $\hat C_P$(SiO$_2$) $= 5.3nR$.  The resulting \DCp is $19.2$ J K$^{-1}$ mol$^{-1}$, or \DCpR$ = 2.3$.  This difference in isobaric heat capacities is driven primarily by the larger value for the SiO$_2$ melt species (SiO$_2$ has a tendency to deviate from the other species). The equilibrium constant for R3 is $K_{\rm eq} = x_{\rm SiO_2} x_{\rm MgO}/x_{\rm MgSiO_3}$ where mole fractions refer to species in the silicate melt and ideal mixing among extant species is assumed.  The value obtained from the thermodynamic data at 6000 K and 1 bar (Appendix) is $0.198$  The value obtained by correcting to 14,000 K using the value for \DCpR of $2.3$ is $0.943$, showing that higher temperatures such as those that would be expected at the core-mantle boundary of the 4$M_\earth$ planet favor MgO + SiO$_2$ over MgSiO$_3$ relative to the speciation at the surface of the magma ocean.  This is an expected result.  However, pressures that would obtain at the core-mantle boundary largely compensate for this effect, as shown next. 

We use the 3rd-order Birch-Murnaghan equation of state for MgSiO$_3$, MgO and SiO$_2$ in the silicate melt based on the compresibilities and their pressure derivatives provided by de Koker and Stixrude (2009), as well as the thermal pressures from that work, to obtain a \DV value of $0.442$ cm$^3$ mol$^{-1}$ for reaction R3 at 14,000K and 450 GPa.  The pressure correction for $-\ln(K_{\rm eq})$ is $\Delta \hat V_{\rm cond, rxn} (\Delta P)/(RT) = 1.7$.  The  equilibrium constant for R3 for $T=14,000$ K and $P=450$ GPa is thus $0.17$, close to the value of $0.2$ obtained at 6000 K and 1 bar.  The net result of correcting the equilibrium constant for reaction R3 for temperature and pressure at core-mantle boundary conditions is effectively no change from the value obtained at $T$ and 1 bar.  For the intra-silicate melt reactions, pressure favors the efficient packing afforded by the larger molecular species (e.g., MgSiO$_3$) over the simple oxides while temperature has the opposite effect, with the result that the equilibrium constants obtained at the surface of the magma ocean are similar to those at core-mantle boundary conditions.      

Pressure corrections are more important if not compensated by temperature effects. For example,  if the pressure correction is applied to the equilibrium constant for R3 at 60 GPa and 6000 K, conditions reflecting possible pressures and temperatures at the surface of the magma oceans, the result is $K_{\rm eq} = 0.06$, or $\ln(K_{\rm eq}) = -2.9$ compared with the 1-bar value of $-1.63$.  

The reactions involving exchange of O and Si between molten silicate and metal exhibit no discernible pressure dependance as reported by \cite{Badro2015}, or contradictory estimates of pressure effects as summarized by \cite{Schaefer2017}.  For internal consistency with the other reactions (see Appendix), we do not apply a pressure correction for the equilibrium constant for reaction R5 describing exchange of H between silicate Fe-rich metal.  The siderophile nature of H appears to increase with $T$ while the pressure dependence of the partitioning is modest, with suggestions of an increase in H in metal relative to silicate between 30 and 60 GPa \citep{Tagawa2021}.  If so, these trends suggest we may be underestimating the H contents of the metal cores in our calculations.  In all events, {\it ab initio} calculations indicate that FeH$_x$ hydrides are stable to hundreds of GPa at relevant temperatures \citep{Bazhanova2012}, and it is reasonable to expect that the thermodynamics of R5 will continue to favor H in metal compared with that in silicate to core-mantle boundary conditions for our 4$M_\earth$ planets. 

We conclude that given the uncertainties and approximations endemic to this exercise, estimates of equilibrium constants at the $T$ and $P$ of the magma ocean surface are adequate to characterize the most basic features of an equilibrated planet.

The effect of the assumption of ideal gas behavior for the H$_2$-rich atmosphere was evaluated using the hydrogen equation of state  reported by \cite{Chabrier2019}.  The compressibility factor, $Z = \rho_{\rm ideal}/\rho = mP/(RT \rho)$ where $\rho_{\rm ideal}$ is the density for the ideal gas and $m$ is the molar mass of H$_2$, varies from 1 to 3 from 0 to 10 GPa at 6000 K (based on the equation of state data provided in tables by Chabrier et al.).  The fugacity coefficient for H$_2$, $\Phi$,  was obtained from the compressibility factor using

\begin{equation}
\ln(\Phi)=\int_0^{P}\frac{Z-1}{P}dP
\label{eqn:z}
\end{equation}
and varies from 1 to 6 over this pressure range.  We were unable to evaluate $\Phi$ beyond this pressure, although we note that $d\Phi/dP$ decreases with P.  While the calculated fugacity coefficients are significant,  the uncertainties in how to treat the solubility of H$_2$ in silicate melt are greater than the effects of non-ideality, justifying the use of ideal gas behavior, as illustrated in Figure \ref{Fig:R15_solubility}.  

\setlength{\abovecaptionskip}{11pt plus 1pt minus 2pt} 

\begin{figure}
\centerline
{\includegraphics[width=0.42\textwidth]{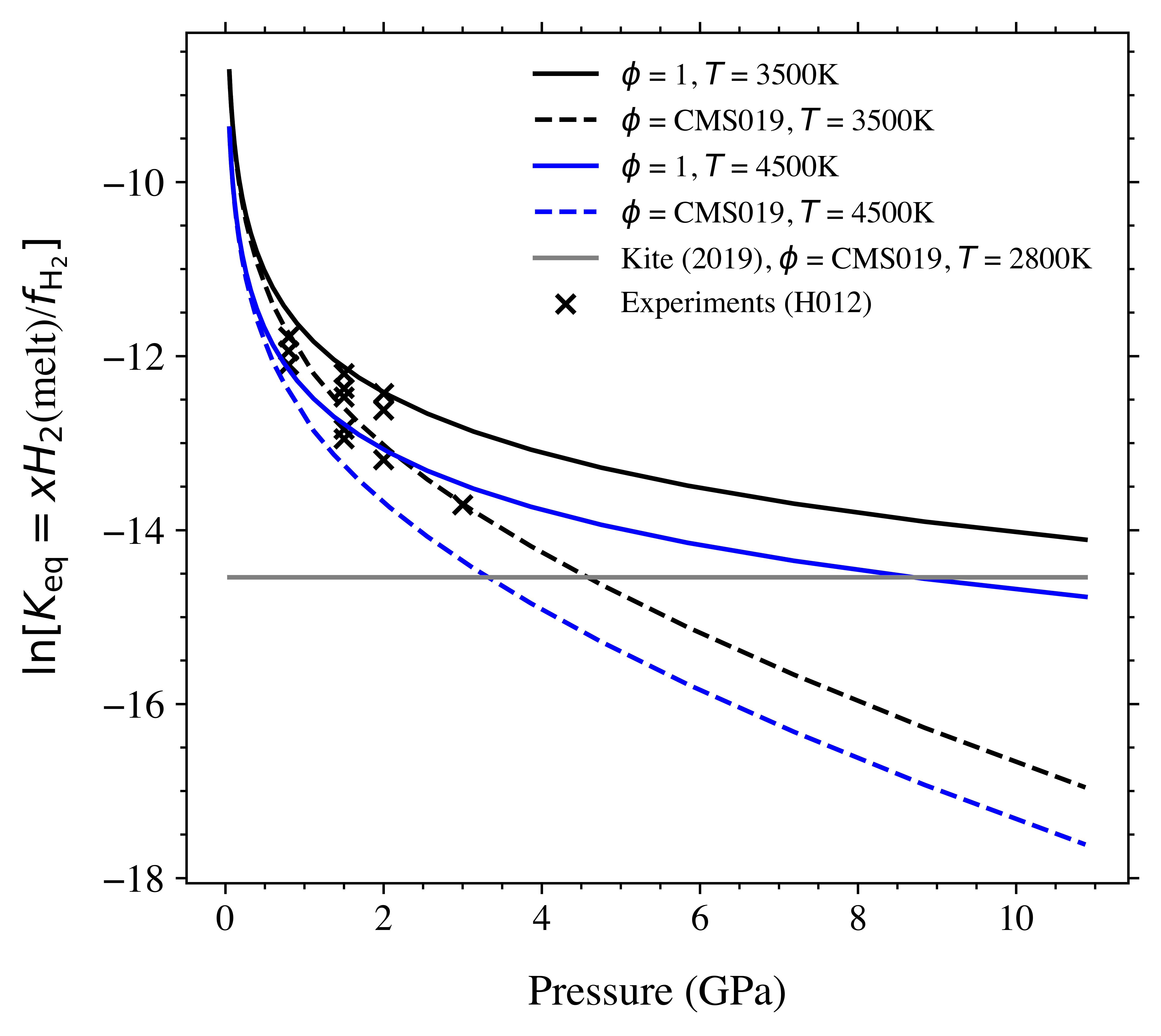}}
\caption{Plot of the equilibrium constant for H$_2$ solubility in molten mafic silicate versus total vapor pressure above the melt (reaction R15).  Crosses are the experimental results from \cite{Hirschmann2012}.  Solid curves show the model used in this work in which the decrease in the equilibrium constant with pressure is attributed solely to the increase in total pressure.  Dashed curves show the same model where the non-ideal behavior of H$_2$ is included based on the equation of state of \cite{Chabrier2019}.  The horizontal line shows the constant equilibrium constant with pressure used by \cite{Kite2019}.  The choice of temperatures for the non-ideal H$_2$ behavior are constrained by the tables provided by \cite{Chabrier2019}.}
\label{Fig:R15_solubility}
\end{figure}

The equilibrium constant for the solubility reaction R15 is $K_{\rm eq} = x_{\rm H_2, silicate}/f_{\rm H_2}$ where $f_{\rm H_2}$ is the fugacity of hydrogen and $f_{\rm H_2}=x_{\rm H_2, gas}\times P\times \Phi$.  Extrapolation of the pressure-dependence of the equilibrium constant for reaction R15 determined by \cite{Hirschmann2012} with experiments up to $\sim 3$ GPa leads to effectively no solubility of H$_2$ beyond 10 GPa, with $x_{\rm H_2, silicate}/x_{\rm H_2, gas}$ values of $2\times 10^{-4}$ and $2\times 10^{-3}$ for the ideal and non-ideal cases at 10 GPa, respectively.  Conversely, if we solve for  $x_{\rm H_2, silicate}/x_{\rm H_2, gas}$ at 3 GPa using the results from \cite{Hirschmann2012}, and assume this distribution coefficient ($K_{\rm D}$) is constant, we obtain a value for $K_{\rm D}$ of $\sim 0.1$ to $\sim 0.3$ from 3500 K to 6000 K based on our internally-consistent thermodynamic parameters for H$_2$ in vapor and H$_2$ in melt (see Appendix).  This provides an intermediate, conservative means  of extrapolating the solubility of H$_2$ in silicate melt to higher pressures.  We find that the experimental data of \cite{Hirschmann2012} are satisfied if the pressure dependence of the equilibrium constant is indeed due entirely to the pressure term in $f_{\rm H_2}$.  In this case, $x_{\rm H_2, silicate}/x_{\rm H_2, gas}$ is constant while $K_{\rm eq}$ decreases, as shown in Figure \ref{Fig:R15_solubility}.  Also shown in Figure \ref{Fig:R15_solubility} is the effect of including the fugacity coefficients from the equation of state for H$_2$ from \cite{Chabrier2019}.  Since the fugacity coefficients, $\phi$, are $> 1$ at higher pressures, our assumption of ideal gas behavior for H$_2$ coupled with a fixed $K_{\rm D}$ leads to an over estimate of the solubility of H$_2$ in melt at pressures greater than about 5 GPa (Figure \ref{Fig:R15_solubility}).  This effect would be exacerbated by assuming a constant equilibrium constant $K_{\rm eq}$  since under those circumstances, higher $f_{\rm H_2}$ forces a higher $x_{\rm H_2, silicate}$.

 Under our model conditions, the atmosphere is beyond the critical point for hydrogen but below the stability of metallic fluid hydrogen as commonly reported \citep[e.g.,][]{Helled2020a}. Our assumption of ideal mixing in the H$_2$-rich fluid phase (ideal gas) is validated by the {\it ab initio} calculations of \cite{Soubiran2015} who find that mixing between H$_2$ and H$_2$O is ideal, with entropy dominating the energetics of mixing, at 2 to 70 GPa and 1000 to 6000 K. 
 
\end{ed}

\section{Results}

Our planets are defined by an initial condition composed of a pure Fe metal, either reactive or unreactive, a silicate mantle with $92.1\%$ (molar) MgSiO$_3$, $3.2\%$ MgO, $3.5\%$ FeSiO$_3$,  $0.7\%$ Na$_2$O , small amounts of SiO$_2$, FeO, and Na$_2$SiO3, and an initial atmosphere  composed of $99.9\%$ H$_2$ gas and $0.1\%$ CO, by mole. In the reactive metal scenario we permit all three phases, metal, silicate mantle, and atmosphere, to reach chemical equilibrium with each other. We refer to this as the ``reactive metal" model. This model applies equally to equilibration of a metal core coalesced at the center of the planet or, alternatively, to equilibration with metal entrained in a vigorously convecting mantle. In the second scenario, we model the same planets, but permit no exchange of material between the metal and the other two phases.  We term this the ``unreactive metal" model and it is meant to simulate the case where the metal is sequestered in a core that is chemically isolated from the rest of the system.

\setlength{\abovecaptionskip}{-11pt plus 1pt minus 2pt} 

\begin{figure*}
\centerline
{\includegraphics[width=0.8\textwidth]{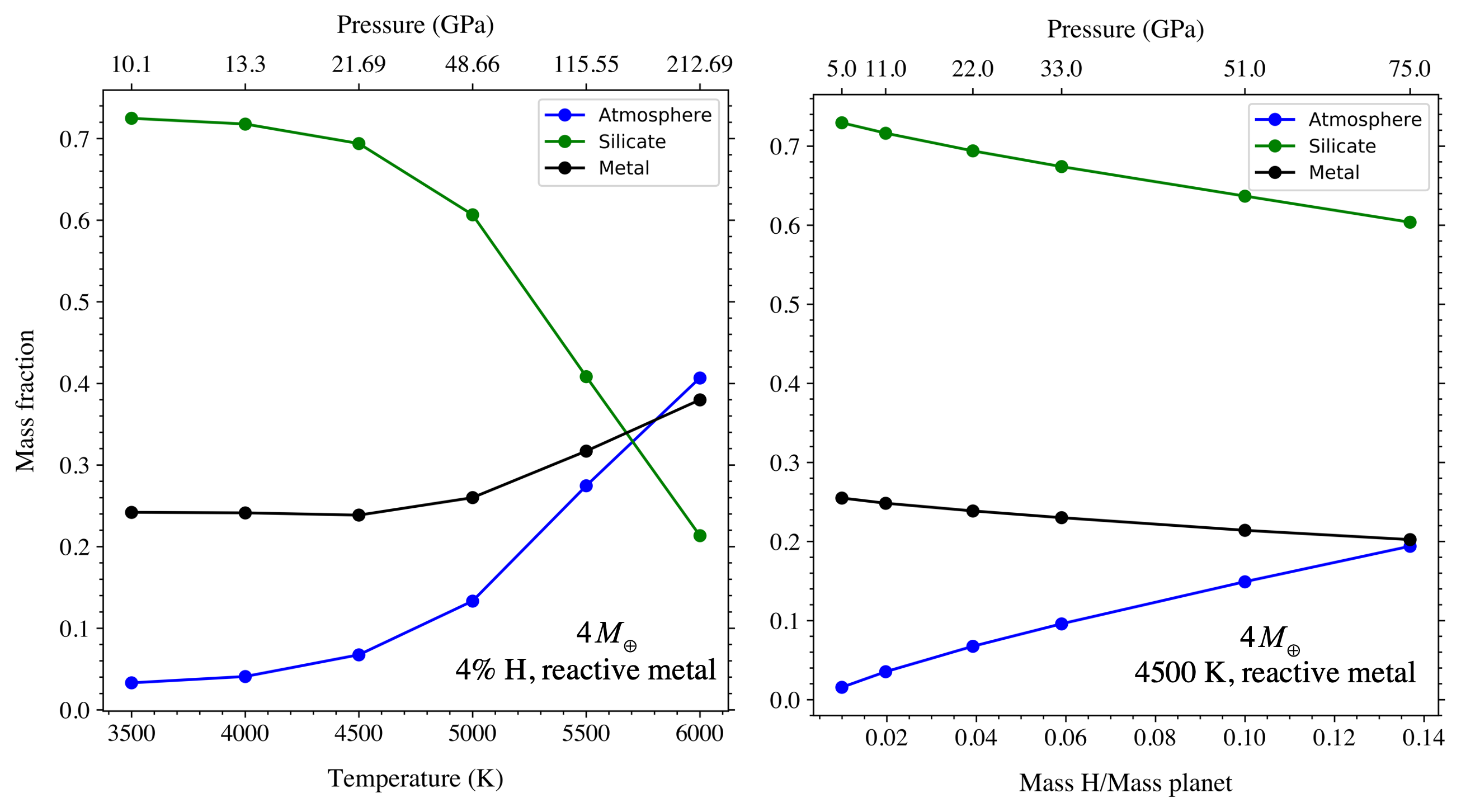}}\hfil
\caption{Calculated mass fractions of atmosphere, silicate, and metal for a 4$M_{\Earth}$ planet with reactive metal. The Panel on the left shows the variability of mass fractions with temperature at the magma ocean - atmosphere boundary with an initial primary atmosphere comprising 4\% of the body by weight.  Pressures along the upper, auxiliary x axis refer to the pressures corresponding to each temperature from the equilibrium solutions.  The panel on the right shows the effect of varying the hydrogen fractions from 1 to 14 \% by weight at a fixed magma ocean surface temperature of 4500 K. Here pressure varies at the constant magma ocean surface temperature of 4500 K. }
\label{Fig:planet_fractions}
\end{figure*}

In both the reactive and unreactive metal scenarios, we assume a planet with a mass of 4$M_{\Earth}$ and vary the mass fraction of the initial H$_2$-rich atmosphere from 1\% to 14\% of the planet's mass, implying that the mass budget is dominated by the metal and silicate portions of the planets. This choice, as well as the planet mass investigated here, are motivated by both observations and modelling of the exoplanet population, which suggests that the most common planet mass prevailing in the close-in super-Earth and sub-Neptune populations is a few Earth masses \citep[e.g.][]{marcy2014a,gupta2020a,rogersj2020a}. In addition, a metal  mass fraction of  25\% is assumed (a conservative mass compared with Earth, for example). We investigated the effects of different temperatures for the magma ocean - atmosphere interface for the case of 4\% H$_2$, with the range a temperatures from 3500 K to 6000 K.  Sub-Neptunes, and by extension the super-Earths that form from them by atmospheric loss, are expected to have initial interior temperatures well in excess of 10,000 K \citep[e.g.][]{ginzburg2016a}. However, as they cool and their envelopes contact, their interior temperatures, including the temperature at atmosphere-mantle boundary, decreases so the temperature of the boundary is case dependent.  We further assume that the mantle and core are close to isothermal. The implications of this assumption were discussed in \S \ref{Thermo} and is revisited below as well.

\begin{ed}
Under our model conditions, the atmosphere is beyond the critical point for hydrogen but below the stability of metallic fluid hydrogen as commonly reported \citep[e.g.,][]{Helled2020a}. Under these conditions, mixing between H$_2$ and H$_2$O is close to ideal (mixing is dominated by the entropy effect with no immiscibility) \citep{Soubiran2015}, justifying in part our assumption of ideal mixing in the atmosphere. We assume the mantle and metallic core are molten throughout. The atmosphere might be better described as an ``envelope" since the boundary between the magma ocean and the supercritical atmosphere may be poorly defined.

\begin{figure*}
\centerline
{\includegraphics[width=0.8\textwidth]{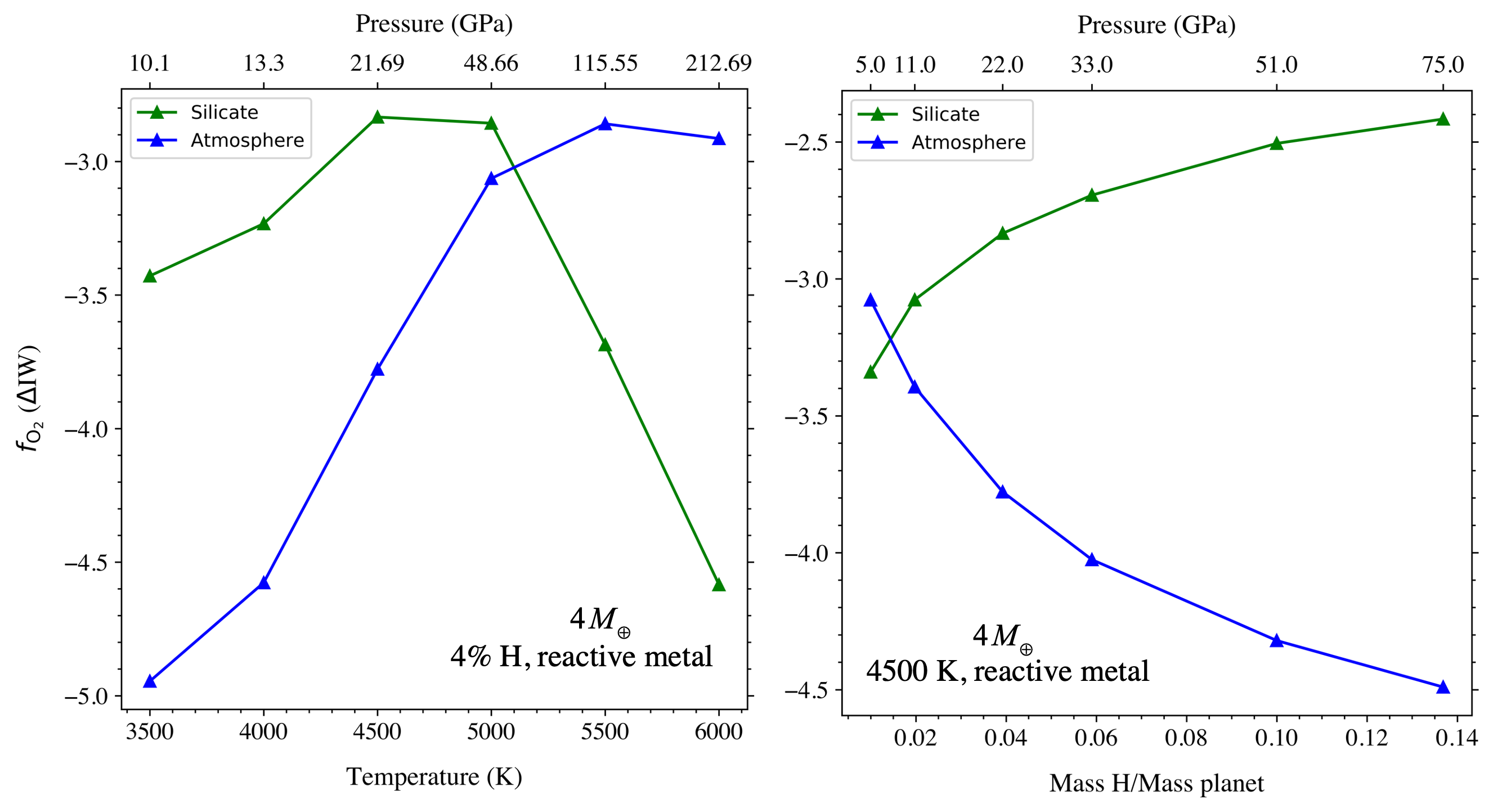}}\hfil
\caption{Calculated oxygen fugacities of the silicate mantle and atmosphere relative to that of the IW buffer for a 4$M_{\Earth}$ planet with reactive metal.  The Panel on the left shows the variability of $f_{\rm O_2}$  with temperature at the magma ocean - atmosphere boundary with an initial primary atmosphere comprising 4\% of the body by weight. Pressures along the upper, auxiliary x axis refer to the pressures corresponding to each temperature from the equilibrium solutions.  The panel on the right shows the effect of varying the hydrogen fractions from 1 to 14 \% by weight at a fixed magma ocean surface temperature of 4500 K.  }
\label{Fig:fO2_reactive}
\end{figure*}

\begin{figure}
\centerline
{\includegraphics[width=0.43\textwidth]{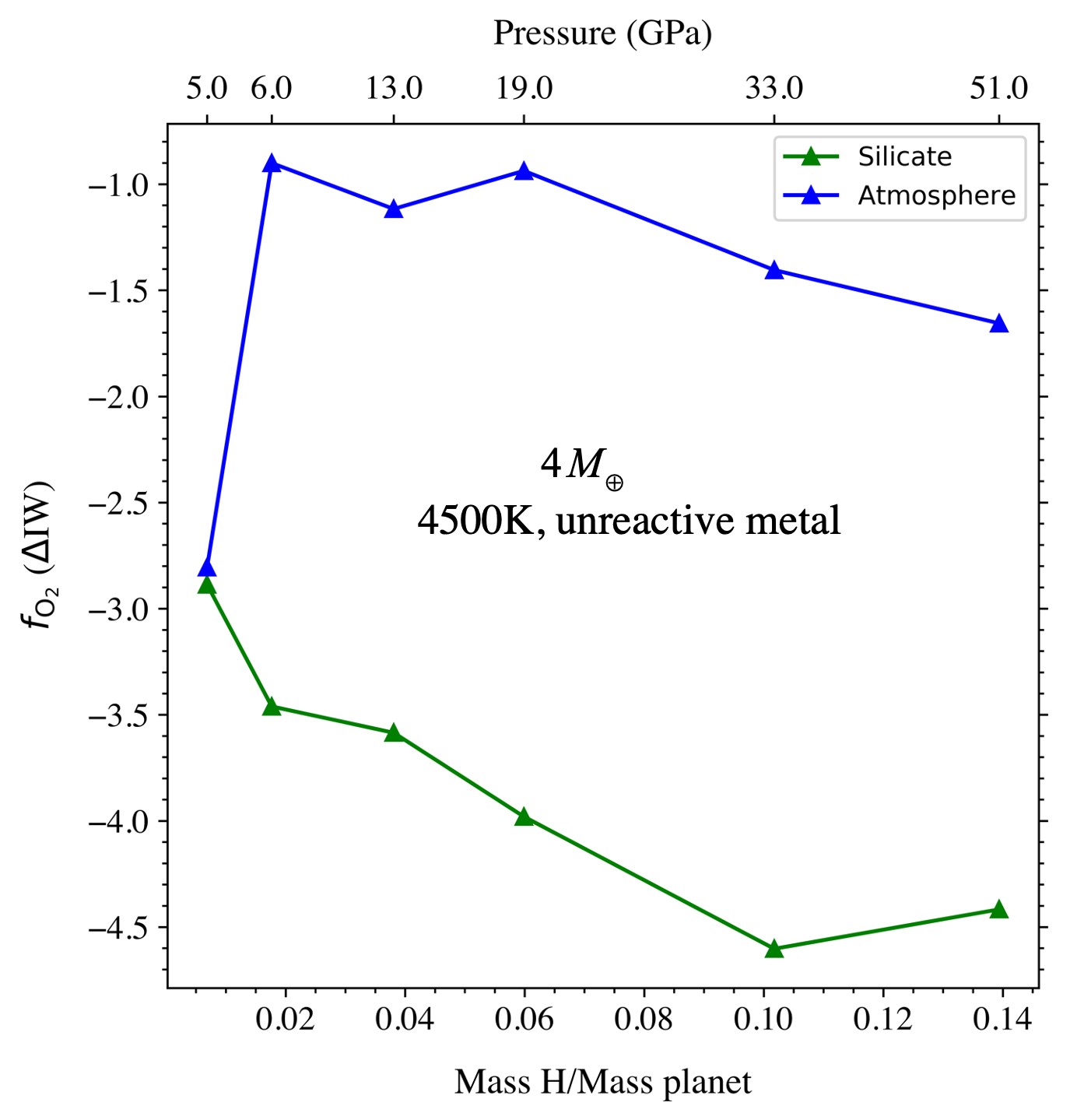}}\hfil
\caption{Calculated oxygen fugacities of the silicate mantle and atmosphere relative to that of the IW buffer for a 4$M_{\Earth}$ planet with unreactive metal.  The effects of varying the hydrogen fractions from 1 to 14 \% by weight at a fixed magma ocean surface temperature of 4500 K are shown here.  }
\label{Fig:fO2_unreactive}
\end{figure}

\subsection{Redox Reactions Between Magma Oceans and Atmospheres}

The overall effect of equilibration of the planets considered is the oxidation of the atmosphere as a result of evaporation of the silicate melt, and reduction of the silicate as a result of ingress of H$_2$ into the melt.  The reactions between silicate and metal are controlled in part by the relative solubilities of H$_2$ and H$_2$O, the latter being a principal product of the reactions.  Since a primary effect of equilibration is an increase in the molecular weight of the atmosphere and a decrease in the mass of silicate, there is a rise in the mass fractions of the atmosphere and metal relative to silicate upon equilibration (Figure \ref{Fig:planet_fractions}).

 We note that O$_2$ is not a likely species in melts, with oxygen atoms comprising the silicate and oxide polymers of the melt likely involved in speciation reactions rather than O$_2$ \citep{Stolper1982}.  However, following convention in geochemistry, we use the simplified thermodynamic expression 

\begin{equation}
\Delta \text{IW} = 2 \text{log} \left( \frac{{x_{\rm FeO}^{\rm silicate}}}{{x_{\rm Fe}^{\rm metal}}}  \right)
  \label{eqn:DIW}
\end{equation}

\noindent to monitor the intrinsic oxygen fugacity of the magma ocean in terms of the decadic logarithm of the oxygen fugacity implied by the oxidation state of Fe. Using Equation \ref{eqn:DIW} we are reporting oxygen fugacities relative to that defined by the reaction between pure Iron and pure FeO (W\"{u}stite), according to the reaction  Fe (Iron) + $1/2$ O$_2$ = FeO (W\"{u}stite). The progress of the overall redox processes is manifest as a decrease in the intrinsic oxygen fugacity in the silicate and metal system relative to that defined by the IW oxygen fugacity buffer.

\begin{figure*}
\centerline
{\includegraphics[width=0.8\textwidth]{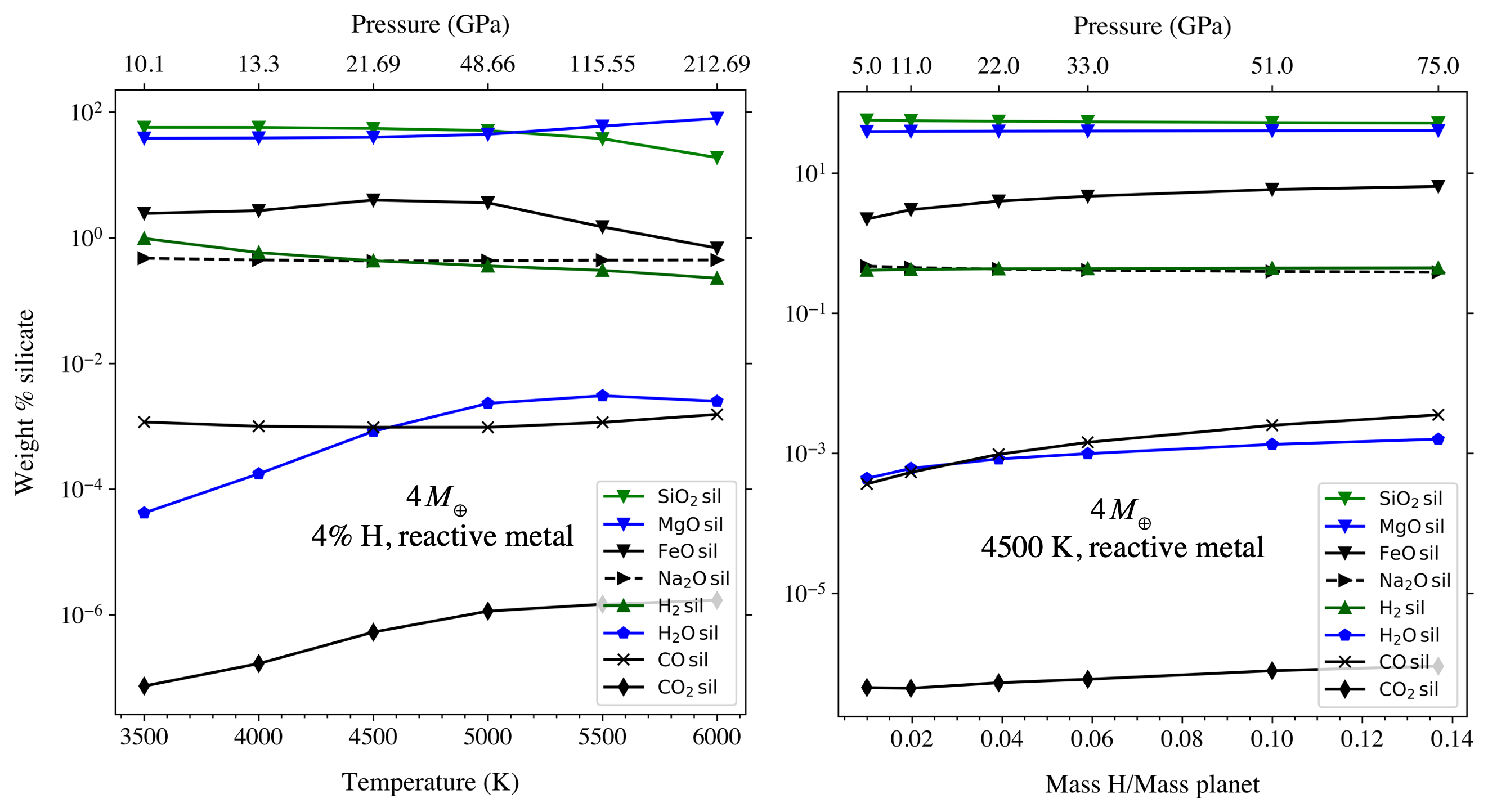}}\hfil
\caption{The chemical composition of the silicate magma ocean equilibrated with the atmosphere for a 4$M_{\Earth}$ planet with reactive metal.  The Panel on the left shows the variability in weight \% with temperature at the magma ocean - atmosphere boundary with an initial primary atmosphere comprising 4\% of the body by weight. Pressures along the upper, auxiliary x axis refer to the pressures corresponding to each temperature from the equilibrium solutions.  The panel on the right shows the effect of varying the hydrogen fractions from 1 to 14 \% by weight at a fixed magma ocean surface temperature of 4500 K.  }
\label{Fig:silicate_reactive}
\end{figure*}

\begin{figure}
\centerline
{\includegraphics[width=0.4\textwidth]{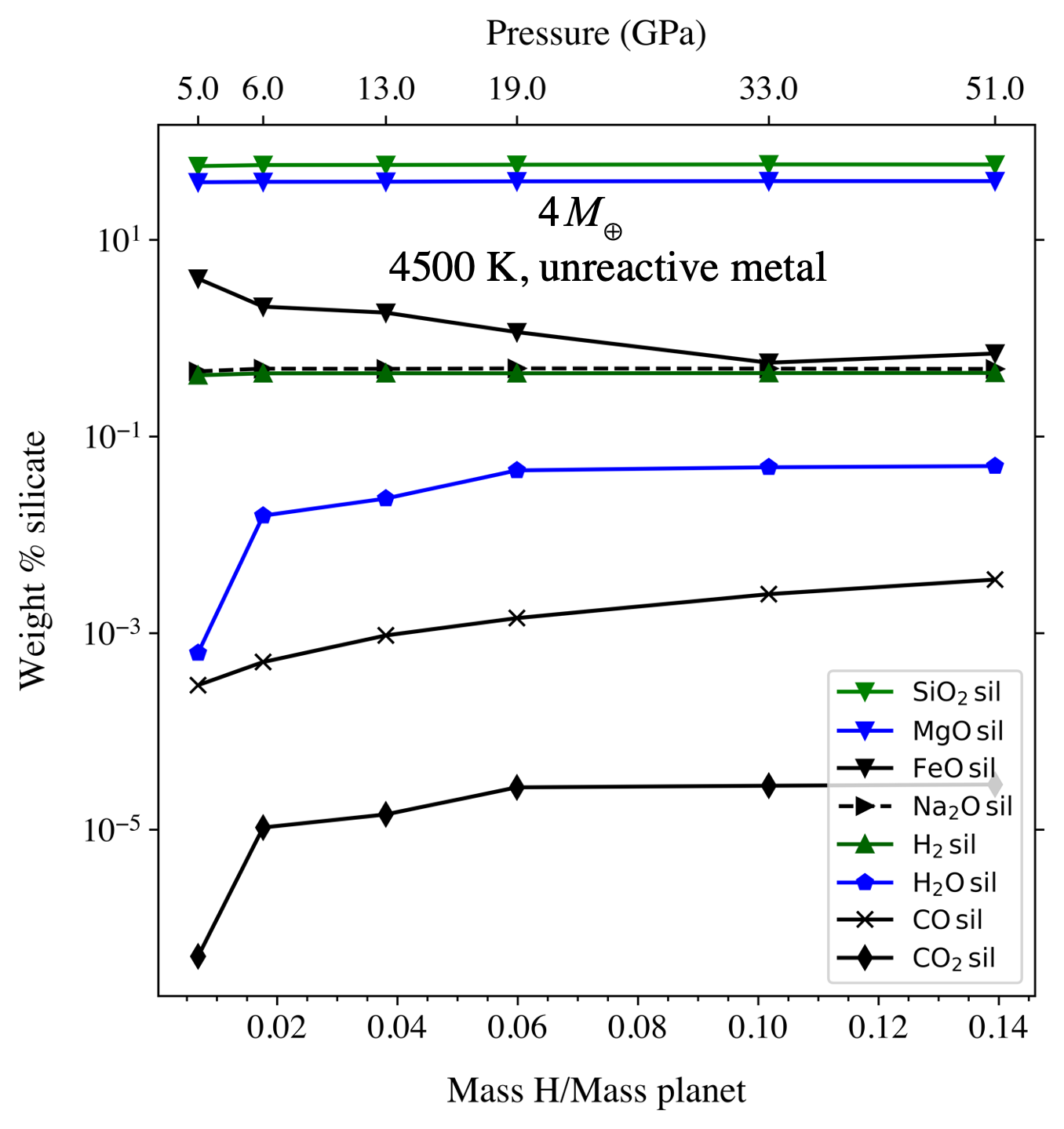}}\hfil
\caption{The chemical composition of the silicate magma ocean equilibrated with the atmosphere for a 4$M_{\Earth}$ planet with unreactive metal.  These results are for a fixed equilibration temperature of 4500 K with the mass fraction of total H varying  from 1 to 14 \% by weight.  }
\label{Fig:silicate_unreactive}
\end{figure}

We apply Equation \ref{eqn:DIW} to our model systems by converting the speciation in the melt to mole fractions of the simple oxides (e.g., by recasting both FeO and FeSiO$_3$ in the melt to ``FeO" in Equation \ref{eqn:DIW}) to be consistent with common usage in the geochemistry and cosmochemistry literature.  The initial condition corresponds to $\Delta$IW values of $-2.8$, similar to, although slightly lower than, that of the Earth.  We use the same reference for the fugacity of oxygen in the atmosphere, where O$_2$ is likely to be an extant species.  We caution, however, that while the oxygen fugacities are reported relative to the same reference, they are not equal at chemical equilibrium because of the absence of O$_2$ as a species in the melt; reporting the oxidation state of the melt in terms of oxygen fugacity relative to the IW buffer is a matter of convention.  

\begin{figure*}
\centerline
{\includegraphics[width=0.8\textwidth]{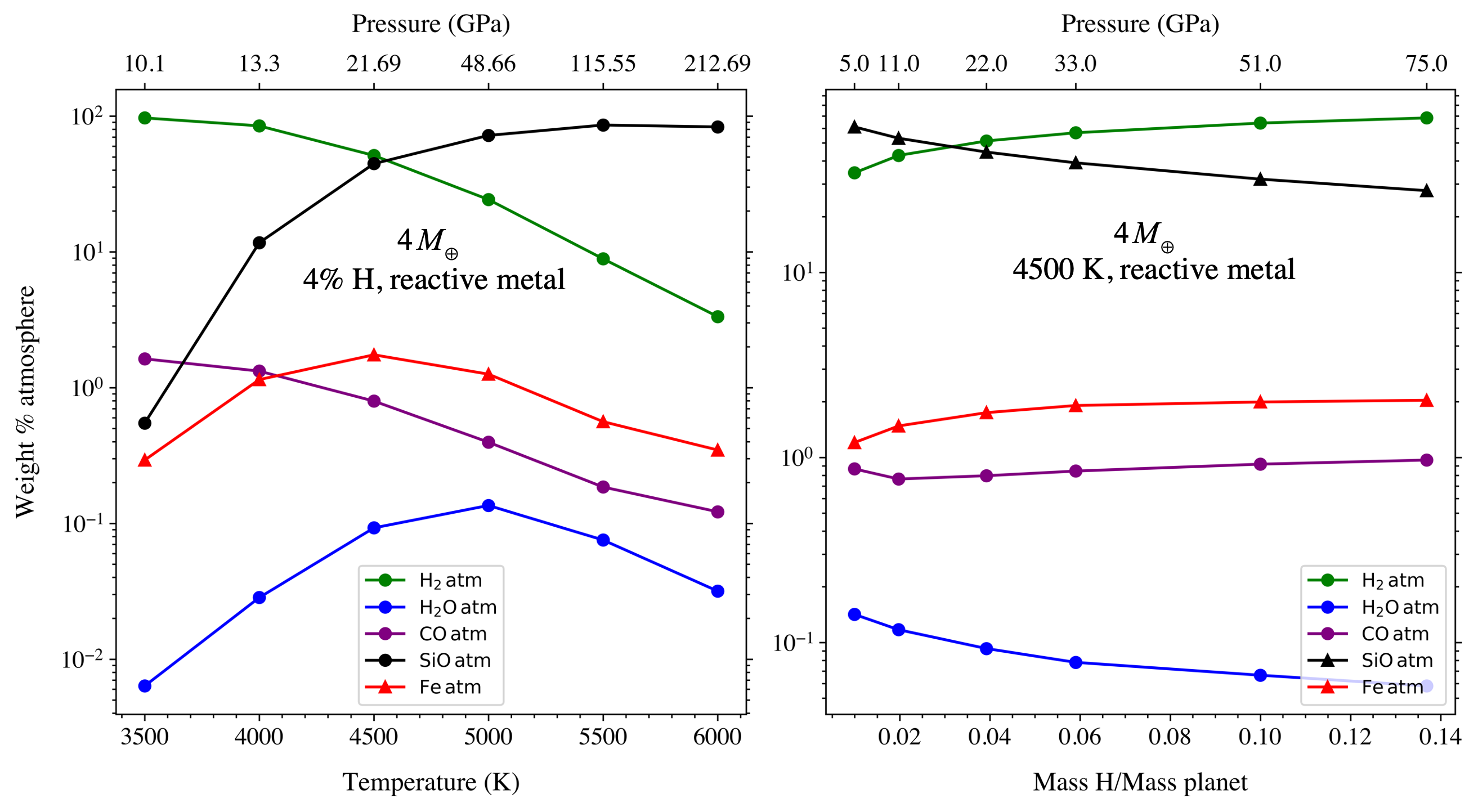}}\hfil
\caption{The chemical composition of an atmosphere equilibrated with the magma ocean for a 4$M_{\Earth}$ planet with reactive metal.  The Panel on the left shows the variability in weight \% with temperature at the magma ocean - atmosphere boundary with an initial primary atmosphere comprising 4\% of the body by weight. Pressures along the upper, auxiliary x axis refer to the pressures corresponding to each temperature from the equilibrium solutions.  The panel on the right shows the effect of varying the hydrogen fractions from 1 to 14 \% by weight at a fixed magma ocean surface temperature of 4500 K.  }
\label{Fig:atmosphere_reactive}
\end{figure*}

The effects of temperature on oxygen fugacity are profound.  At temperatures above about 5000 K, evaporation releases large fractions of the silicate mantle into the atmosphere or its supercritical equivalent.  The nature of the resulting silicate-H mixture will have to be characterized using minimization of free energies, and is beyond the scope of this exercise.  Nonetheless, at more modest magma ocean temperatures of  $\le 4500$ K, one expects to find relatively reduced silicate for the reactive metal case (Figure \ref{Fig:fO2_reactive}) and atmospheres that are orders of magnitude more oxidized than the initial primary atmosphere assumed here.  At most conditions, the atmosphere remains more reduced than the silicate in the reactive metal cases.  This is reversed for the cases where metal is unavailable for reaction (Figure \ref{Fig:fO2_unreactive}).  Under these circumstances, evaporation oxidizes the atmosphere and there is no metal sink for H$_2$ dissolved in the silicate, so the atmosphere has  $\Delta \text{IW}$ values of $\sim -1.0$ to $-1.5$ while silicate has $\Delta \text{IW}<-3.5$ (Figure \ref{Fig:fO2_unreactive}).

The exceptions to these trends occur where the primary atmosphere of H$_2$ comprises less than 1\% by mass of the planet.  In those cases, most of the atmosphere  is dissolved into the silicate, and what remains has oxygen fugacities similar to the condensed planet.  

The calculated speciation in the silicate melts is shown in Figures \ref{Fig:silicate_reactive} and \ref{Fig:silicate_unreactive} for the reactive and unreactive metal cases, respectively.  Among the volatiles, the relative concentrations are H$_2$ $>>$ H$_2$O $\sim$ CO $>>$ CO$_2$ for the reactive metal case.  In the unreactive metal case, the lack of a metal sink for hydrogen results in higher concentrations of H$_2$O (discussed below).  In all cases, the concentration of H$_2$ is $<$ 1\% by weight and H$_2$O is $\le 0.1$ wt\%. 

The effects of oxidation of the atmosphere are shown by examining the speciation in the atmosphere as a function of both temperature and the mass of the initial primary atmosphere.  For the case of hydrogen comprising 4\% by mass of the planet, equilibration between the condensed planet and the overlying atmosphere results in an increasing mass fraction of SiO and H$_2$O in the atmosphere with increasing temperature (Figure \ref{Fig:atmosphere_reactive}).  At a temperature of about 4500 K the mass fractions of SiO and H$_2$ are equal.  Beyond this temperature, the dominant species in the atmosphere is SiO.  Higher fractions of total H result in the continued dominance of H$_2$, but SiO remains an abundant species (Figure \ref{Fig:atmosphere_reactive}). 

The evaporation reactions depend critically on the concentrations of the simple oxides FeO, MgO, SiO$_2$, and Na$_2$O in the silicate melts.  Speciation reactions R1, R3, and R6 strongly influence these concentrations by defining the activities of the oxides in the melts. The activity coefficients implied by the calculated concentrations of these oxides are obtained by taking the ratio of these concentrations to the values implied by the bulk chemistry of the melts.  For example, $\gamma_{\rm MgO} = x_{\rm MgO}$/[MgO] where $x_{\rm MgO}$ is the mole fraction of MgO in the melt obtained from the thermodynamic calculations and [MgO] is the molar concentration obtained by recasting the bulk composition of the melt into moles of oxides. At 4500 K our results indicate activity coefficients of approximately $0.40$ to $0.35$ for MgO and SiO$_2$, respectively, near unity for FeO, and $0.02$ for Na$_2$O. 

\subsection{Water Budget}

\setlength{\abovecaptionskip}{11pt plus 1pt minus 2pt}
\begin{figure}
\centerline
{\includegraphics[width=0.4\textwidth]{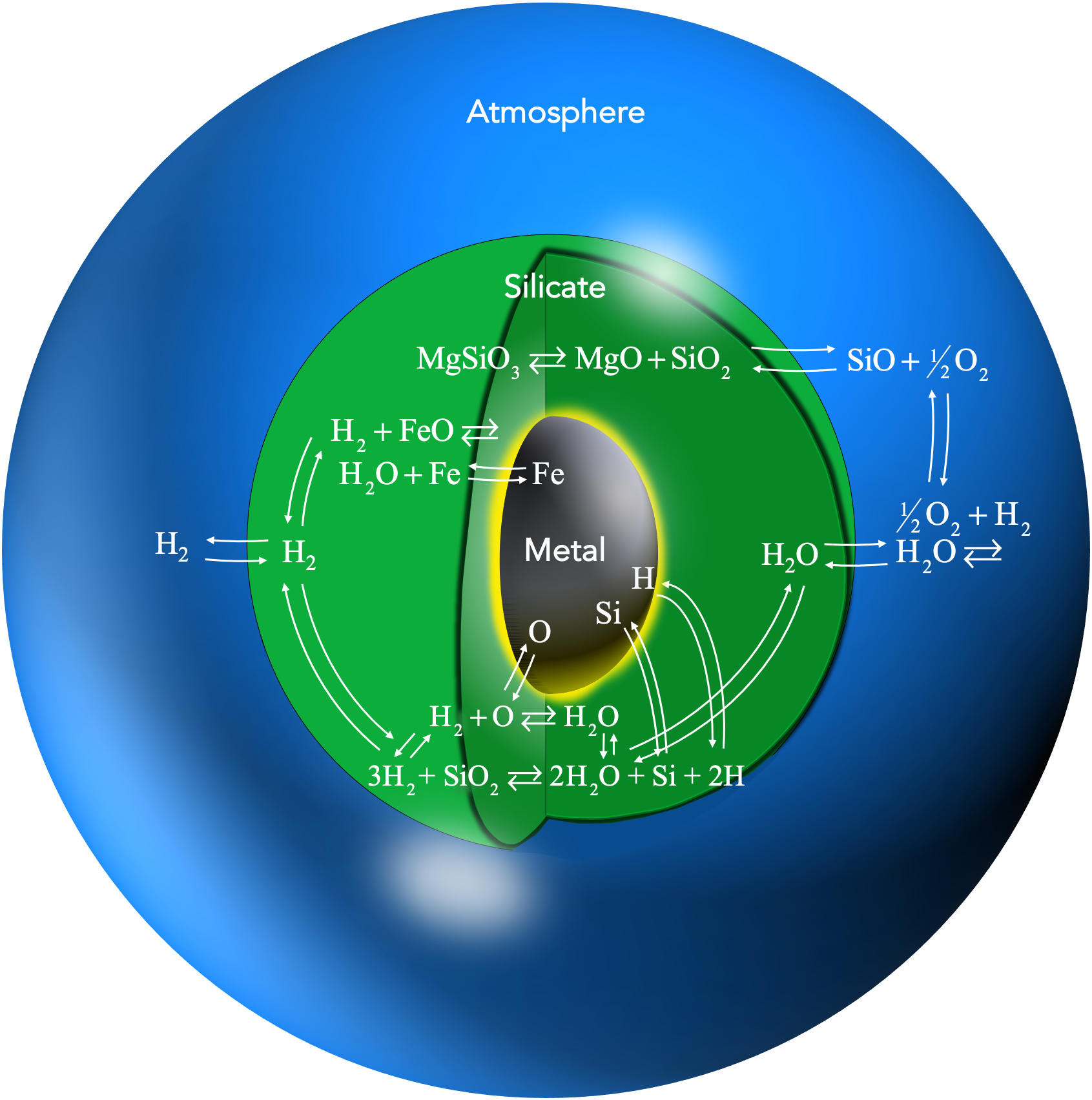}}
\caption{Simplified depiction of the net reactions comprising the hydrogen cycle for a planet with an H$_2$-rich primary atmosphere in the reactive core model. Redox chemistry leads to water formation in the atmosphere and light elements in the metal core. Fe, H, O, and Si refer to the elements in the metal phase.  }
\label{fig:water_cycle}
\end{figure}

Water is perhaps the most important product of the reaction between the planetary ``core" and the overlying atmosphere.  A common error is to interpret equilibria such as R1 through R18 as reaction pathways in the literal sense.  They are not.  Rather, as we emphasize above, they are basis vectors spanning reaction space, and the actual reaction pathways are merely described by this basis.  Nonetheless, it is instructive to use linear combinations of these reactions to describe not only the equilibrium state but the actual reactions that lead to equilibrium as evidenced by the net transfer of chemical constituents between the atmosphere, silicate melt, and metal melt upon reaching equilibrium. In the case of the production of water, one can isolate two pathways for water production in these models that are revealed by comparing the reactive core and unreactive core calculations.  Where metal is participating in the chemistry, intra-melt reactions that lead to the formation of water begin with H$_2$ dissolution into the silicate melt by reaction R15 (Figure \ref{fig:water_cycle}). Dissolved hydrogen donates electrons to the silicate melt, and the reduction of the melt can be described by the combination of reactions R7 and R5, leading to the net reaction

\begin{rxn}
\begin{array}{r}
3{\rm H}_{2,{\rm silicate}} + {\rm SiO}_{2,{\rm silicate}}\rightleftharpoons 2{\rm H}_2{\rm O}_{\rm silicate} + \\ {\rm Si}_{\rm metal} + 2{\rm H}_{\rm metal},
\end{array}
\label{eqn:H2Oproduction}
\end{rxn}

\noindent where the production of water in the silicate is seen to be the biproduct of the redox reaction that drives hydrogen into the metal. Water made in this way effuses from the silicate to the atmosphere and vice versa as prescribed by the equilibrium constant for reaction R16.  The general stoichiometry of this reaction is verified by positive correlation between H$_2$O in the silicate melt and Si and H in metal in the reactive metal scenario (not shown). 

\setlength{\abovecaptionskip}{-1pt plus 1pt minus 2pt}
\begin{figure}
\centerline
{\includegraphics[width=0.45\textwidth]{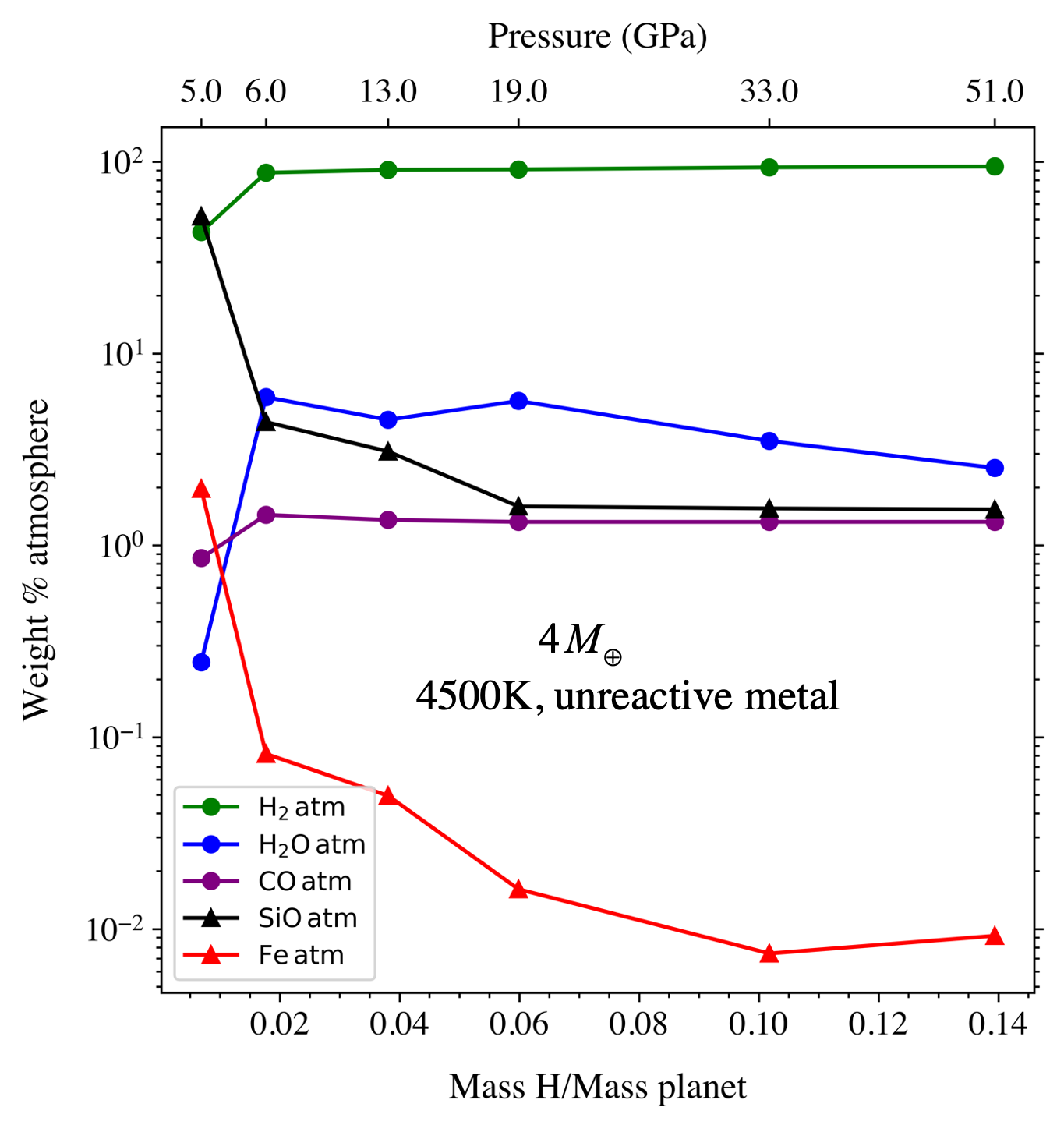}}\hfil
\caption{The chemical composition of an atmosphere equilibrated with the magma ocean for a 4$M_{\Earth}$ planet with no reactive metal as a function of magma ocean surface pressure.  Each pressure represents a different total mass fraction of hydrogen, ranging from 1\% to 14\% at a fixed magma ocean surface temperature of 4500 K.  }
\label{Fig:atmosphere_unreactive}
\end{figure}

\begin{figure*}
\centerline
{\includegraphics[width=0.8\textwidth]{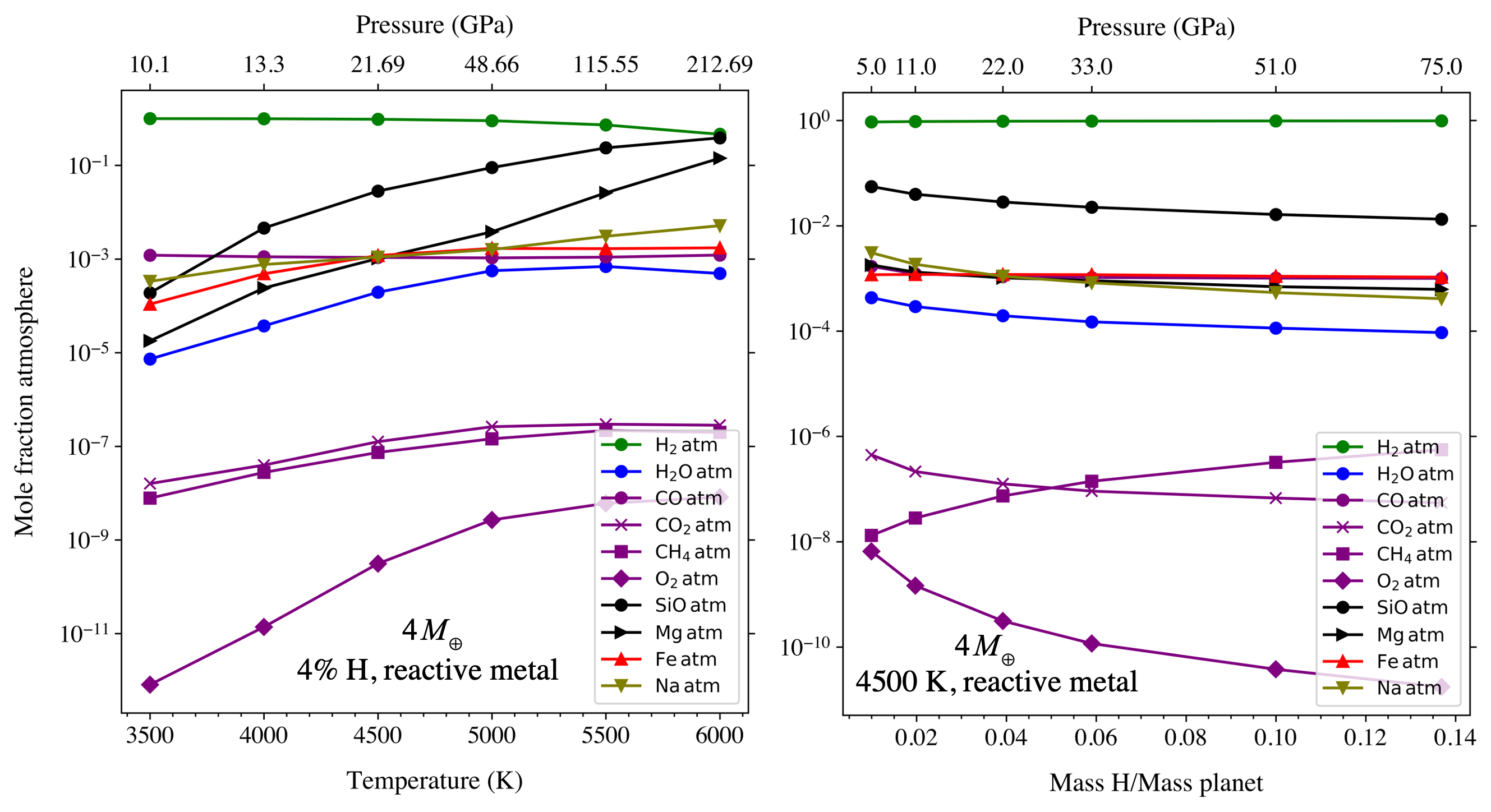}}\hfil
\caption{Mole fractions of species in the atmosphere as a function of  magma ocean - atmosphere boundary temperature with an initial H$_2$-rich primary atmosphere comprising 4\% of the body by weight. These results are for the reactive metal scenario. Pressures along the upper, auxiliary x axis refer to the pressures corresponding to each temperature from the equilibrium solutions.  The panel on the right shows the effect of varying the hydrogen fractions from 1 to 14 \% by weight at a fixed magma ocean surface temperature of 4500 K. }
\label{Fig:atm_molefracs}
\end{figure*}
However, this is not the sole source of water. Significant amounts of water are produced even where the metal core is unreactive (Figure \ref{Fig:atmosphere_unreactive}).  The ultimate source of water in those cases is evaporation of SiO and, importantly, O$_2$ (R13) that oxidizes the atmosphere. Oxidation drives the formation of water by reaction R10.  The H$_2$O produced in the atmosphere (Figure \ref{Fig:atmosphere_unreactive}) equilibrates with the magma ocean below, delivering water to the melt, and with no metal to serve as a sink for H$_2$O in the melt by redox reactions, water builds up to wt \% levels. When metal is not  accessible for equilibration (unreactive metal model), the concentrations of H$_2$O in the silicate melts are greater than those with metal by factors of about 50 to 100$\times$ (Figure \ref{Fig:silicate_unreactive}) and are controlled simply by the exchange equilibrium between the atmosphere and the silicate melt (R16). 

While the detailed descriptions will vary with different choices for species and reaction basis sets, the overall effect of dissolution of H$_2$ resulting in reduction of silicate melt, oxidation of the atmosphere, and production of water, is an important consequence  of bringing hydrogen into contact with silicate melt at relevant temperatures. The production of water by reactions between H$_2$ atmospheres and the condensed portions of the planet has been described previously with an emphasis on the role of iron metal \citep[e.g.][]{Ikoma2006,Kite2021}. Here we have shown that water will be produced at equilibrium even where metal is not reacting. Water should not be scarce on a planet in which a primary H$_2$ atmosphere reacts with a silicate magma ocean. 

\begin{figure*}
\centerline
{\includegraphics[width=0.43\textwidth]{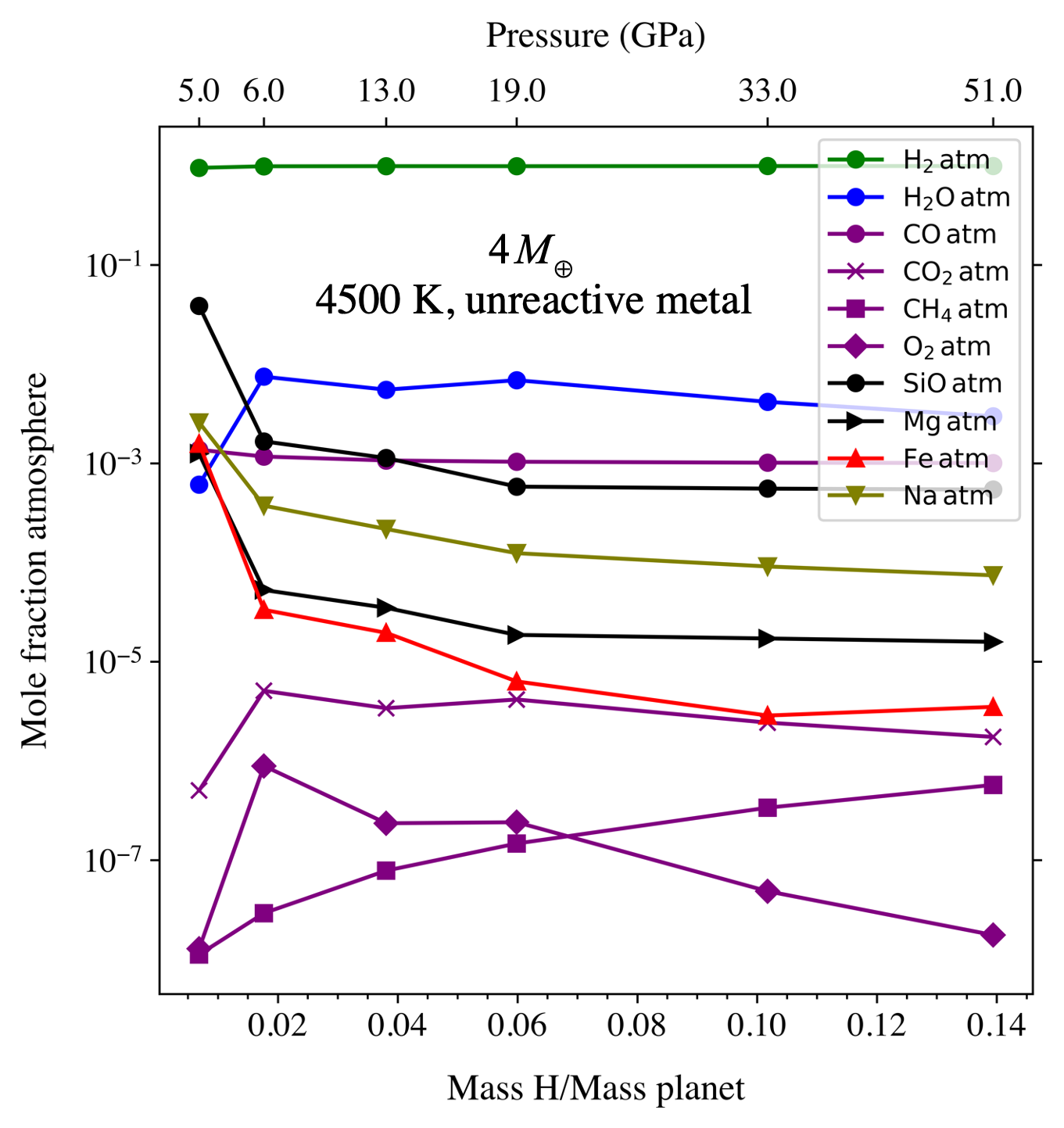}}\hfil
\caption{Mole fractions of species in the atmosphere as a function of  magma ocean - atmosphere boundary temperature with an initial H$_2$-rich primary atmosphere comprising 4\% of the body by weight.  These results are for the unreactive metal scenario. Pressures along the upper, auxiliary x axis refer to the pressures corresponding to each temperature from the equilibrium solutions.  The panel on the right shows the effect of varying the hydrogen fractions from 1 to 14 \% by weight at a fixed magma ocean surface temperature of 4500 K. }
\label{Fig:atm_molefracs_unreactive}
\end{figure*}

\subsection{Detailed Chemistry of the Atmosphere}\label{chemat}

In addition to the major species H$_2$ and SiO, our calculated bulk atmospheres are characterized by increasing concentrations of Mg and Fe with increasing temperature of the magma ocean surface, as well as slight increases in CO$_2$,  CH$_4$, and O$_2$ (Figure \ref{Fig:atm_molefracs}).  The effect of more massive primary atmospheres is to simply dilute these species with greater abundances of H$_2$.  Below 4500 K we predict atmospheres with compositions with relative abundances in terms of molar concentrations similar to H$_2$ $>$  SiO $>$ CO $\sim$ Na $\sim$ Mg $>$ H$_2$O $>>$ CO$_2$ $\sim$ CH$_4$ $>>$ O$_2$ for the reactive metal case (Figure \ref{Fig:atm_molefracs}).  For the unreactive metal case, at a magma ocean surface temperature of 4500 K, the atmosphere differs from the reactive metal case in that H$_2$O becomes the second most abundant species on a molar basis, after H$_2$ (Figure \ref{Fig:atm_molefracs_unreactive}).

\subsection{Light Elements in Metal Cores}

In the reactive metal model, the presence of metal regulates the amount of water in the magma ocean.  There is a strong correlation between the mole fraction of H$_2$O in silicate melt and the mole fraction of O in metal, for example.  More broadly, where metal is present, the reactions in our models produce water mediated by a steady-state amount of the light elements H, O, and Si in the metal.  Greater masses of hydrogen in the primary atmosphere result in more water production that in turn results in a metal core with weight percent levels of H and up to tens of weight percent of O (Figure \ref{Fig:metal_weightpercent}).

At the more modest temperatures of reaction, large amounts of H and O in the metal modify the density of the metal significantly.  We interpret our results to mean that provided that hydrogen can be efficiently transported from the atmosphere through the silicate mantle and into the metal core, we expect these metal cores of exoplanets with hydrogen rich envelopes to have hydrogen and oxygen in significant abundances.   The presence of significant quantities of H in metal cores of planets has been suggested previously by, for example, \citet{Hirao2004} and \cite{Terasaki2009}, and most recently on the basis of experimental partitioning data by \citet{Tagawa2021} and {\it ab initio} calculations for hydrogen partitioning between silicate and metal \citep{Li2020_1038}.  

Transfer of H to metal is sufficiently efficacious that H in metal is predicted to be among the most important reservoirs for hydrogen. For example, for the 4\% total H case,  H in metal is predicted to be the third most important reservoir for H on the planet at equilibration temperatures of less than 4500 K. Metal H replaces H$_2$ in silicate melt as the second most important hydrogen reservoir for equilibration at temperatures $>$ 4500 K (Figure \ref{Fig:Hfraction}).  The remainder of the H is stored mainly in H$_2$ in the silicate and H$_2$O in the atmosphere.  In the unreactive metal scenario, H$_2$ in silicate melt and H$_2$O in the atmosphere are second only to H$_2$ in the atmosphere in the fraction of total H.  The exception is where the total H is $\le$ 1\%, where total pressure at the surface of the magma ocean is $\sim$ 5 GPa.  In this case, H$_2$O comprises less than 1/100th of the total H budget (Figure \ref{Fig:Hfraction_unreactive}), and virtually all H is present as H$_2$ in the atmosphere and in silicate melt. 

At extreme pressures and temperatures the phase represented by a ``metal" composed of Fe with $\sim 33$ wt\% O and $\sim 6$ wt\% H, as produced in our calculations at temperatures of 6000 K and above, will need to be assessed with phase equilibria.  Using the current T-dependent calibrations for the reactions that deliver light elements to the core, even more extreme compositions would be predicted.  For example, if we run our calculations for a magma ocean surface temperature of 4500 K but set the temperature for reactions R2, R4, R5, and R7 to 10,000 K (to represent the higher T of a core-mantle boundary), the resulting ``metal" core has the stoichiometry Fe$_{0.6}$SiO$_{2}$H$_{5.5}$. This Si, O and H-rich material is likely to retain a metallic character at the high pressures of the core \citep{Scipioni2017}. 

\begin{figure*}
\centerline
{\includegraphics[width=0.8\textwidth]{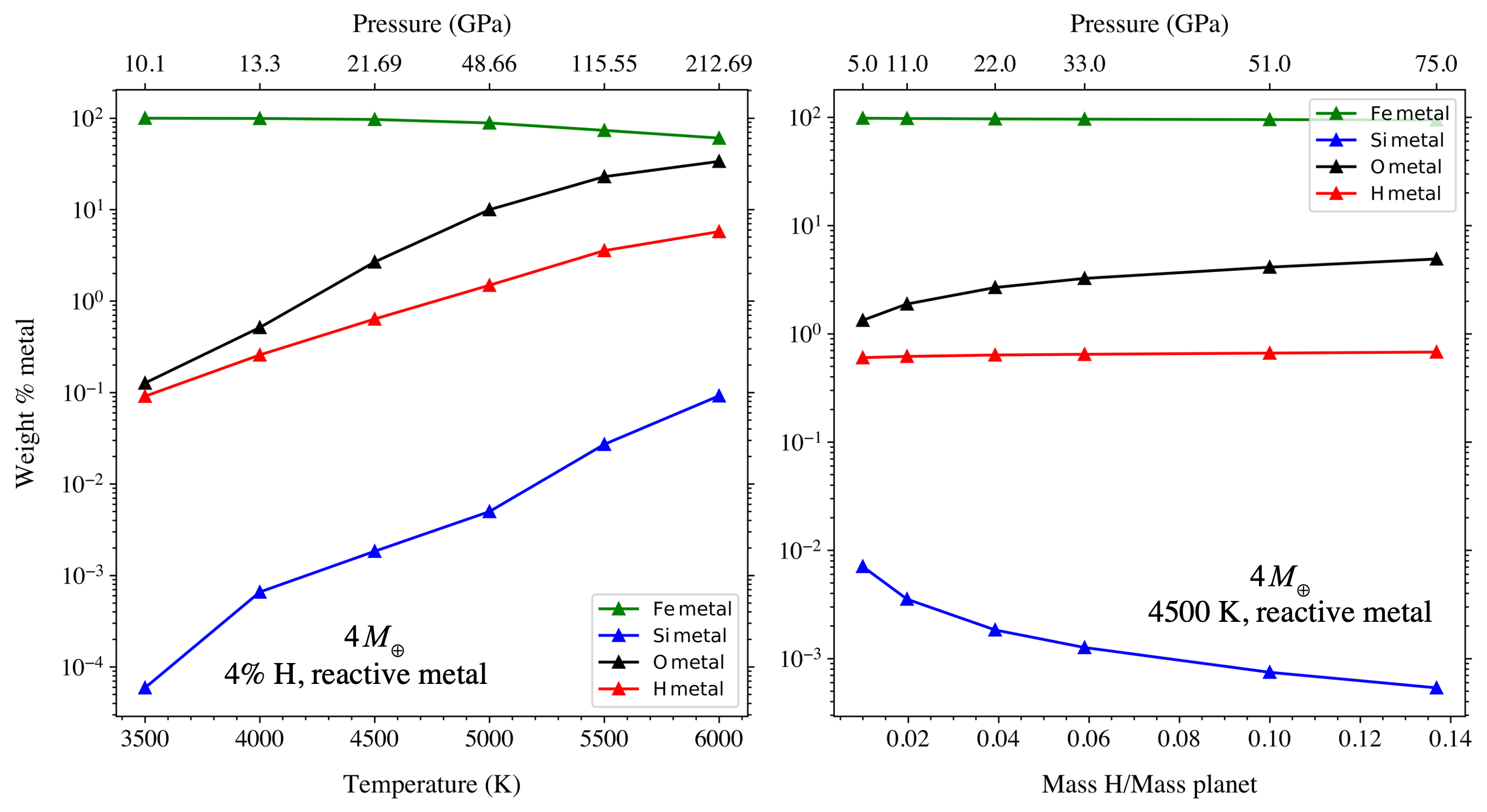}}\hfil
\caption{Weight \% elements in metal for the reactive core model for a 4$M_{\Earth}$ planet.  The Panel on the left shows the variability in weight \% with temperature at the magma ocean - atmosphere boundary with an initial primary atmosphere comprising 4\% of the body by weight. Pressures along the upper, auxiliary x axis refer to the pressures corresponding to each temperature from the equilibrium solutions.  The panel on the right shows the effect of varying the hydrogen fractions from 1 to 14 \% by weight at a fixed magma ocean surface temperature of 4500 K.  }
\label{Fig:metal_weightpercent}
\end{figure*}

\begin{figure*}
\centerline
{\includegraphics[width=0.8\textwidth]{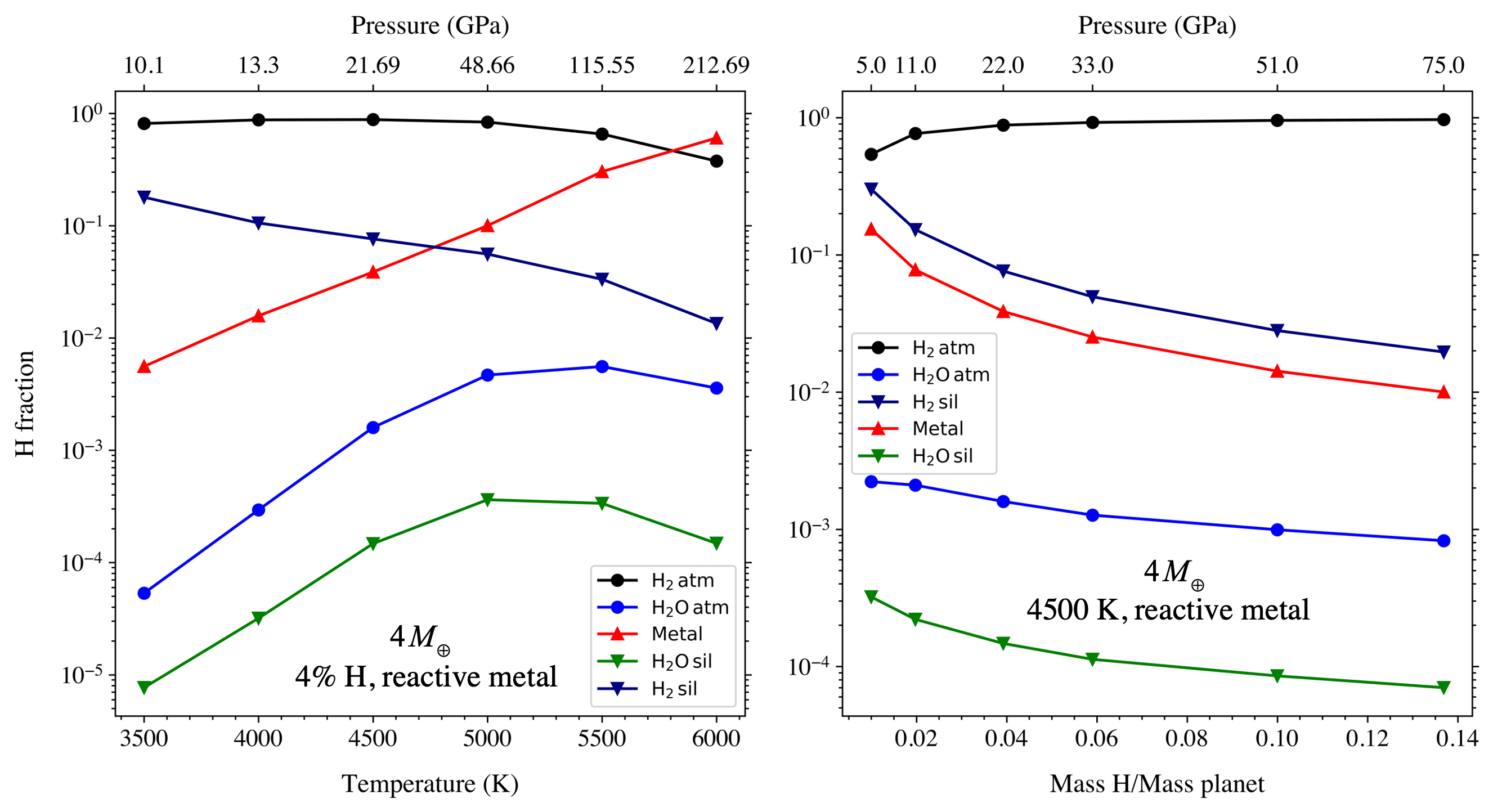}}\hfil
\caption{The distribution of H in an equilibrated  4$M_{\Earth}$ planet.  The Panel on the left shows the variability in the fraction of H in each reservoir as a function of  temperature at the magma ocean - atmosphere boundary with an initial primary atmosphere comprising 4\% of the body by weight. Pressures along the upper, auxiliary x axis refer to the pressures corresponding to each temperature from the equilibrium solutions.  The panel on the right shows the effect of varying the hydrogen fractions from 1 to 14 \% by weight at a fixed magma ocean surface temperature of 4500 K.  Here, H fraction refers to the fraction of total H present in each of the reservoirs.}
\label{Fig:Hfraction}
\end{figure*}

\begin{figure}
\centerline
{\includegraphics[width=0.4\textwidth]{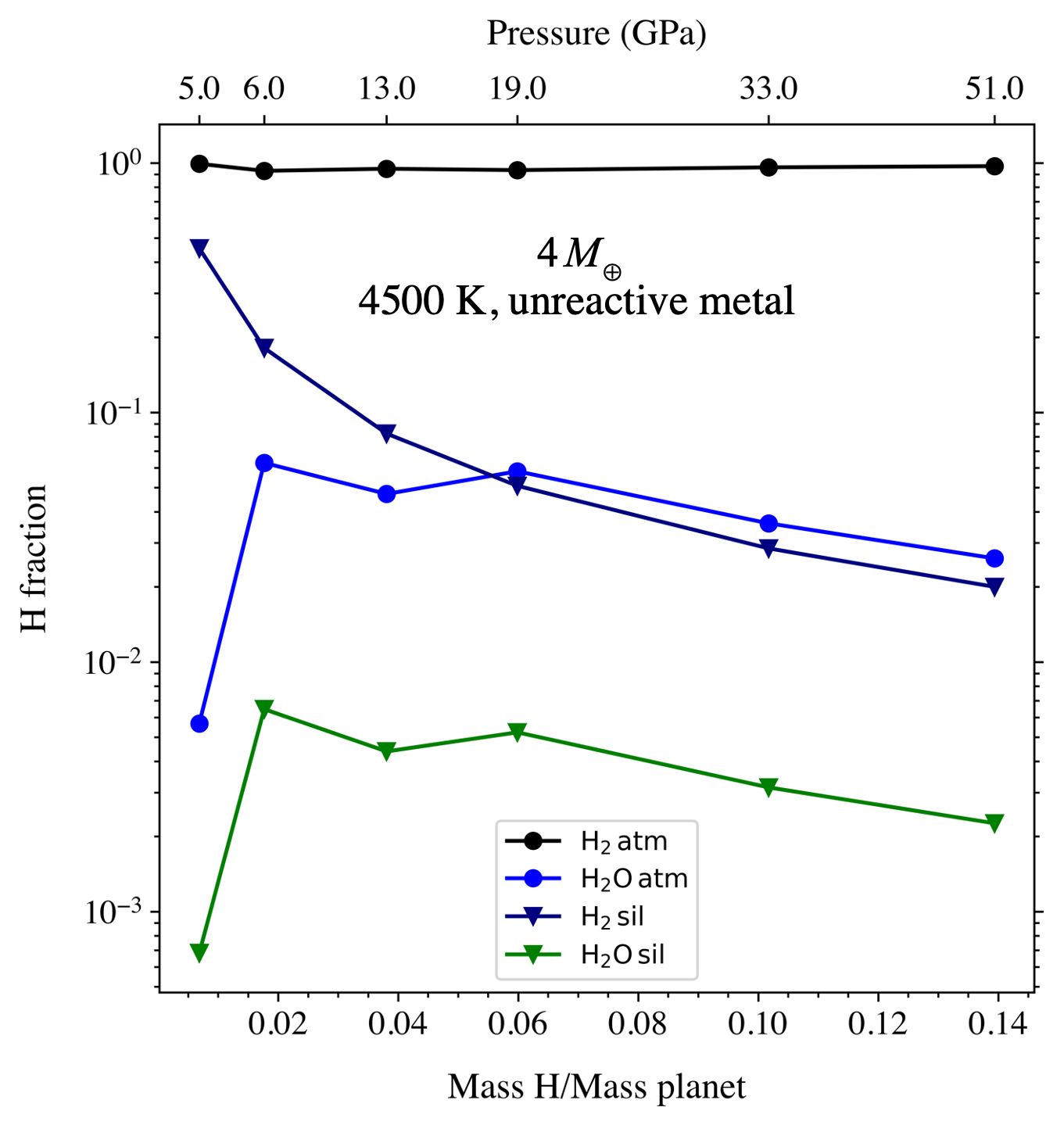}}\hfil
\caption{The distribution of H in an equilibrated 4$M_{\Earth}$ planet with unreactive metal as a function of varying the hydrogen fractions from 1 to 14 \% by weight at a fixed magma ocean surface temperature of 4500 K.  Here, H fraction refers to the fraction of total H present in each of the reservoirs.}
\label{Fig:Hfraction_unreactive}
\end{figure}

\end{ed}

\section{Observational Consequences}

\subsection{Mass-Radius Relation of Super-Earths}\label{s4}

\begin{ed}
The large fractions of H and O sequestered in metallic cores of planets hosting a primary, H$_2$-rich atmosphere, would have significant consequences for our interpretation of mass-radius relationships for these planets because of the resulting decreases in core densities.
We calculate the reduction in metal (e.g., core) density resulting from addition of H and O to Fe compared to pure Fe, expressed in terms of the uncompressed density. We use the {\it ab initio} results of \cite{Huang2019} to include the effects of O on density, and we use the experimental results summarized by \cite{Li2019} for the effects of H. The effect of O is to decrease $\rho_0$ by $1.2\%$ per weight $\%$ O in Fe metal.  The effect of H is to decrease $\rho_0$ by $8.7\%$ per weight $\%$ H in Fe metal.   The results suggest that uncompressed density deficits of up to approximately 12\% are expected at relatively low equilibration temperatures of 4500 K, and  that density deficits of more than 50\% may be expected for higher temperatures such as those that may obtain at the boundary between the silicate magma ocean and a metallic core (Figure \ref{Fig:density_deficit}).

\begin{figure*}
\centerline
{\includegraphics[width=0.8\textwidth]{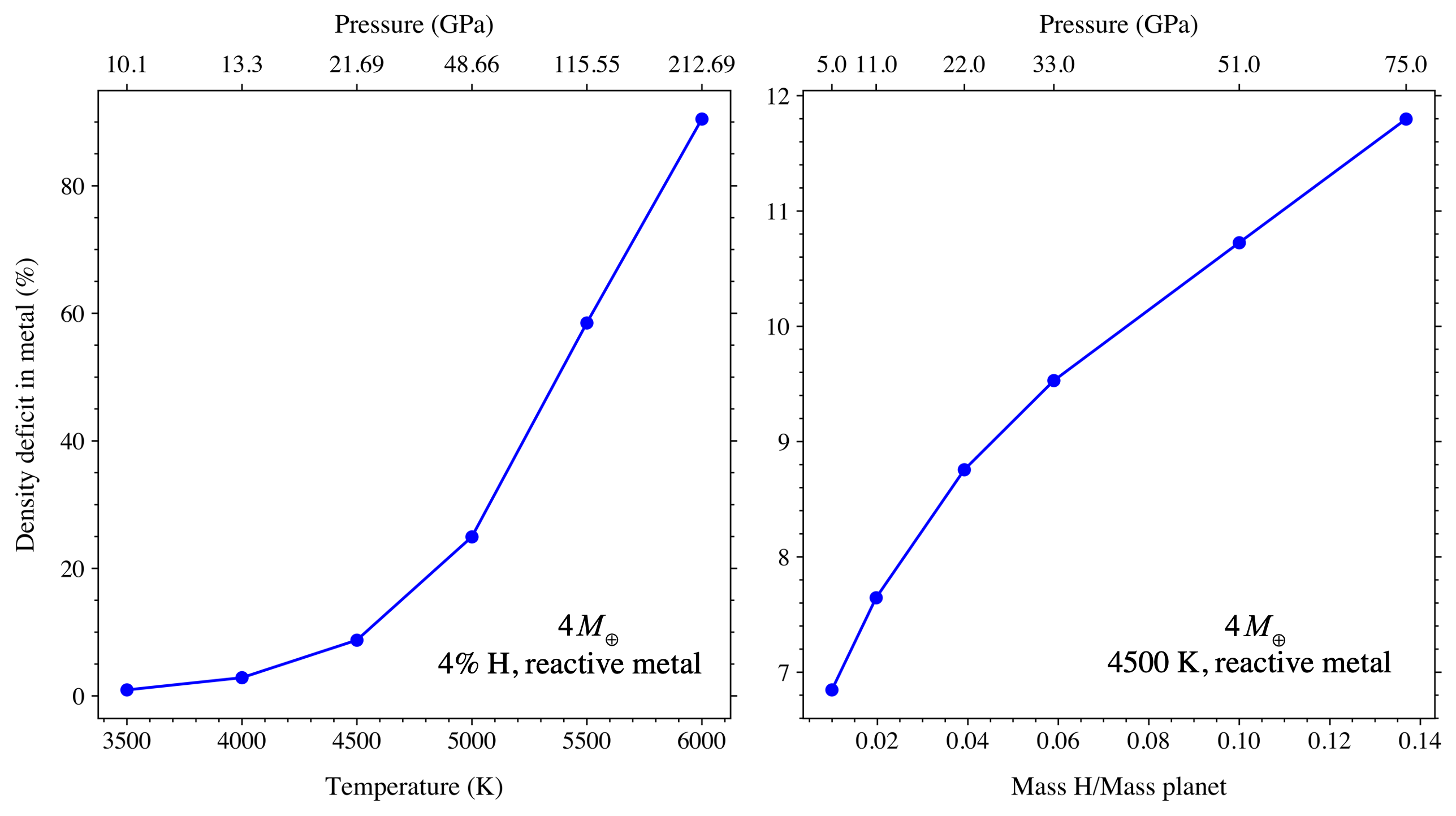}}\hfil
\caption{The deficit in uncompressed density, $\rho_0$, for a  4$M_{\Earth}$ planet as a function of temperature of equilibration and total H.  The Panel on the left shows the variability in $\rho_0$ as a function of  temperature at the magma ocean - atmosphere boundary with an initial primary atmosphere comprising 4\% of the body by weight. Pressures along the upper, auxiliary x axis refer to the pressures corresponding to each temperature from the equilibrium solutions.  The panel on the right shows the effect of varying the hydrogen fractions from 1 to 14 \% by weight at a fixed magma ocean surface temperature of 4500 K.  }
\label{Fig:density_deficit}
\end{figure*}

Hydrogen may be a principal light element in Earth's core.  \cite{Umemoto2015} showed that $1\%$ by mass H in the liquid outer core is consistent with seismological constraints and would account for most of the $10\%$ density deficit in the core \citep[e.g.,][]{Birch1964}. Therefore, as a reality check, we tested our model to see if it can reproduce the proposed effects of H, as well as O, in Earths core. We calculated the density deficit in the metal core of a proto-Earth with a mass of  $0.9\, M_\Earth$ and a primary H$_2$ atmosphere comprising 0.05\% by mass of the planet \citep[e.g.,]{ginzburg2016a}. We use a magma ocean surface temperature of 2600 K and a metal core - silicate mantle equilibration temperature of 3000 K, consistent with Earth's metal-silicate equilibration temperature suggested by trace element partitioning \citep{Wood2008}. Under these conditions, our reaction network predicts $1.08\%$ H and $0.61\%$ O by mass in the core and a mass deficit in the core of 10\%, consistent with observations.  Most of the atmosphere is consumed by the mantle and core and what remains is composed of roughly equal parts H$_2$ and CO on a molar basis, with rest mainly composed of  H$_2$O.   

\end{ed}

However, whether or not H can reach the metal core of a fully differentiated planet depends on several factors, including the transport mechanism through the silicates and the timescales for differentiation of the silicate mantle and metal core, if it happens at all. For example, it may be challenging to transport hydrogen into the metal core, after differentiation, since any significant amounts of hydrogen absorbed by the upper mantle would make it more buoyant and hence likely hinder large scale convective mixing, and thus also hindering inward transport towards the metallic core.

It is therefore instructive to turn the question around and ask, for rocky planets massive enough to have hosted a primary, H$_2$-rich atmosphere, what are the observational consequences of several wt\% of H in a iron metal core of super-Earths?  For those planets that may have once had H$_2$-rich atmospheres but subsequently lost them, what limits do current exoplanet data provide with respect to the effects of these atmospheres on the planet cores? In order to answer this question, we  constructed simple planetary interior models of  differentiated planets.  We use these models to examine the consequences of H and O in the metal core for the mass-radius relationship. The mass-radius relations are obtained in the usual way \citep[e.g.][]{seager2007a} by solving for the mass, $m(r)$, contained within a given radius, $r$,
\begin{equation}\label{emr1}
    \frac{d m(r)}{d r}=4\pi r^2 \rho(r)
\end{equation}
together with the requirement for hydrostatic equilibrium 
\begin{equation}\label{emr2}
    \frac{dP(r)}{dr}=-\frac{G m(r) \rho(r)}{r^2},
\end{equation}
and the appropriate equation of state (EOS)
\begin{equation}
   P(r)=f(\rho(r),T(r)),
\end{equation}
where $f$ is a function relating density and pressure. $P(r)$, $\rho(r)$, and $T(r)$ are the pressure, density and temperature of the planet, and $G$ is the gravitational constant.

\begin{figure}
\centering
\includegraphics[width=0.5\textwidth]{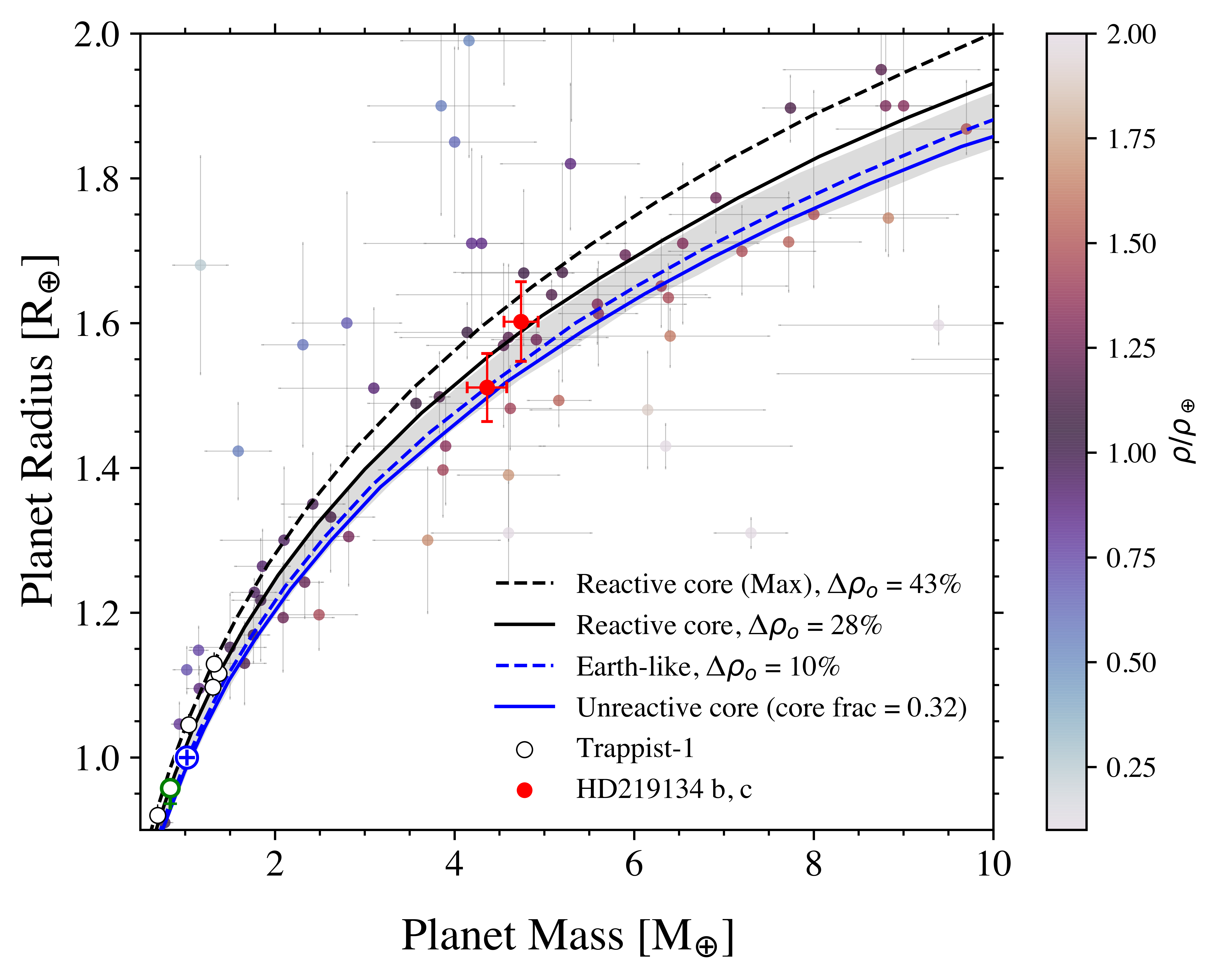}
\caption{Mass-Radius relations for differentiated rocky planets.  The solid blue curve is for an Earth-like planet with a pure Fe core and a MgSiO$_3$ perovskite mantle.  This case corresponds to an unreactive core in our calculations. All curves correspond to a core mass fraction of 32\%. The solid and dashed  black curves represent solutions for planets with reactive cores, in which the Fe metal cores have density deficits of 28\% and 43\% due to H and O in the metal. A 28\% density deficit in Fe corresponds to about 2\% H and 8\% O by mass, corresponding to a metal core - silicate mantle equilibration temperature of about 5000 K with 4 wt\% total H (black line) in our models. The 43\% density deficit in Fe represents the maximum amount of H and O we expect to be dissolved in Fe, because otherwise the core will become less dense than the mantle, and corresponds to 3\% H and 14\%O by mass (black dashed line). Also shown are masses and radii for planets where uncertainties in mass and radius are  $< 35\%$ and $< 15\%$, respectively, as reported in the NASA Exoplanet Archive. Each datum is  color-coded for nominal densities relative to Earth (see text for details). The grey shaded area spans the region for pure Fe core mass fractions ranging from $0.17$ to $0.34$, representing an estimate of the $1\sigma$ density uncertainty among rocky super-Earths derived from planetary evolution and mass-loss models that successfully explain the radius gap in the exoplanet size distribution \citep{rogers2020b}.  Solar-system rocky planets are shown for reference.
} 
\label{Fig:M_vs_R}
\end{figure}

For a given EOS of interest, we integrate equations (\ref{emr1}) and (\ref{emr2}) from the planet's center, using the inner boundary condition $m(0)=0$ and $P(0)=P_{\rm center}$. The outer boundary condition is given by $P(R_{\rm p})=0$.  For the metal-silicate differentiated planets we switch from one EOS to the other while maintaining continuity in pressure across the core-mantle boundary. The planet's mass, $M_{\rm p}$ and radius, $R_{\rm p}$, are uniquely determined by $P(0)=P_{\rm center}$ and $P(R_{\rm p})=0$ for a specified mass fraction for the metal core. Numerically integrating equations (\ref{emr1}) and (\ref{emr2}) for core and mantle over a range of central pressures, and thus planet masses, yields the mass-radius relationships shown in Figure \ref{Fig:M_vs_R}. 

To examine the effects of the density of the core on the mass-radius relationship, we started with a base state that is Earth-like, consisting of a pure Fe metal core containing $32\%$ of the total mass and with the remainder of the planet composed of MgSiO$_3$ perovskite. A known difficulty with simple two-layer models of this sort is that, by using the EOS for the high-pressure form of solid Fe in the hexagonal close-packed structure,  Fe($\epsilon$), the radius of Earth at 1 $M_\Earth$ is underestimated.  Earth's core is under dense relative to pure Fe as a result of the presence of light elements \citep[e.g.,][]{Birch1964}.  However, even allowing for the known $\sim 10\%$ deficit in density relative to pure iron, the use of solid Fe($\epsilon$) results in an underestimate of Earth's radius.  The reason is that a molten iron core is less dense than the solid.  \cite{Zeng2016} fit the Preliminary Reference Earth Model (PREM) to arrive at suitable equations of state for Earth's liquid outer core and lower mantle. However, the EOS derived for the liquid metal {\it a priori} includes the density effects of light elements in Earth's core.  We sought to construct an internally consistent model by using the experimental data for the density of molten Fe at high temperatures and pressures to derive an uncompressed density for liquid Fe to be used with a Vinet EOS \citep[e.g.,][]{seager2007a,Anderson2001}.  Extrapolation of the data of \cite{Kuwayama2020} yields an uncompressed density, $\rho_0$, of $7.2 \pm 0.1$ g/cm$^3$. This is slightly greater than the PREM liquid outer core value of $7.05$ and considerably less than the Fe($\epsilon$) value of 8.4 \citep[e.g.,][]{Smith2018}, as expected.  For the silicate we use the Mg-perovskite third-order Birch-Murnagham EOS of \cite{Karki2000} described by \cite{seager2007a}.  None of our EOS include variations with temperature, and we do not include variations in $\mathrm{Fe}/(\mathrm{Fe}+\mathrm{Mg})$ in the mantle. We note that the small mass fractions of H$_2$ and H$_2$O in the silicate predicted by our calculations correspond to density deficits of $\le 0.1 \%$ for the reactive core model.

\setlength{\abovecaptionskip}{11pt plus 1pt minus 2pt}

We model the effects of H and O in metal cores by incorporating our calculated zero-pressure densities, $\rho_0$, into the equations of state for the core. 
The solid blue curve in Figure \ref{Fig:M_vs_R} is our base state with a pure molten Fe core and a MgSiO$_3$ perovskite mantle. Adding a $10\%$ density deficit to this core due to the presence of light elements, as evidenced for Earth \citep{Birch1964, Badro2015}, shifts the curve to align well with Earth, validating our model for the influence of light elements in metal on the mass-radius relation.

Figure \ref{Fig:M_vs_R} displays the mass-radius relation for various relevant density deficits. As expected, density deficits in the metal increase the radii of the cores. A planet with an Earth-like metallic core mass fraction, but with a $28\%$ density deficit in the metal would be the result of  2 wt\% dissolved H and 8 wt\% dissolved O, corresponding to a metal core - silicate mantle equilibration temperature of about 5000 K with 4 wt\% total H (black line) in our models. The core radius  is about $60\%$ of the total radius in the absence of any atmosphere (shown as black line in Figure \ref{Fig:M_vs_R}). 

While thermodynamics permits higher fractions of H in the metal under the right circumstances, we note that a physical limit is imposed by the requirement that the metal core density needs to be greater than, or equal to, that of the mantle at the core-mantle boundary.  This limit corresponds to a density deficit for the core of about $43\%$ for the equations of state used here. The corresponding mass-radius relation for maximum core-mass-density deficit is shown as the dashed black curve in Figure \ref{Fig:M_vs_R}, where we assumed Earth-like core mass fractions. The black-dashed line is for all practical purposes identical to that of pure MgSiO$_3$ perovskite. This maximum density deficit of $43\%$ corresponds to about $5\%$ by mass H in the metal core, or a combination of $3\%$  H and $14\%$ O by mass. 

In order to illustrate the applicability of our calculations, we extracted planet masses and radii with uncertainties of $< 35\%$ and $< 15\%$, respectively, from the NASA Exoplanet Archive. We color-coded the planets with masses $< 10 M_{\Earth}$ to derive nominal densities relative to Earth using $M_{\rm p}/M_\Earth =(R_{\rm p}/R_\Earth)^{4}(\rho_{\rm p}/\rho_\Earth)^{4/3}$ where we make use of the relationship between mass and radius for rocky planets from \cite{valencia2006a} (Figure \ref{Fig:M_vs_R}). 

\begin{figure}
\centering
\includegraphics[width=0.45\textwidth]{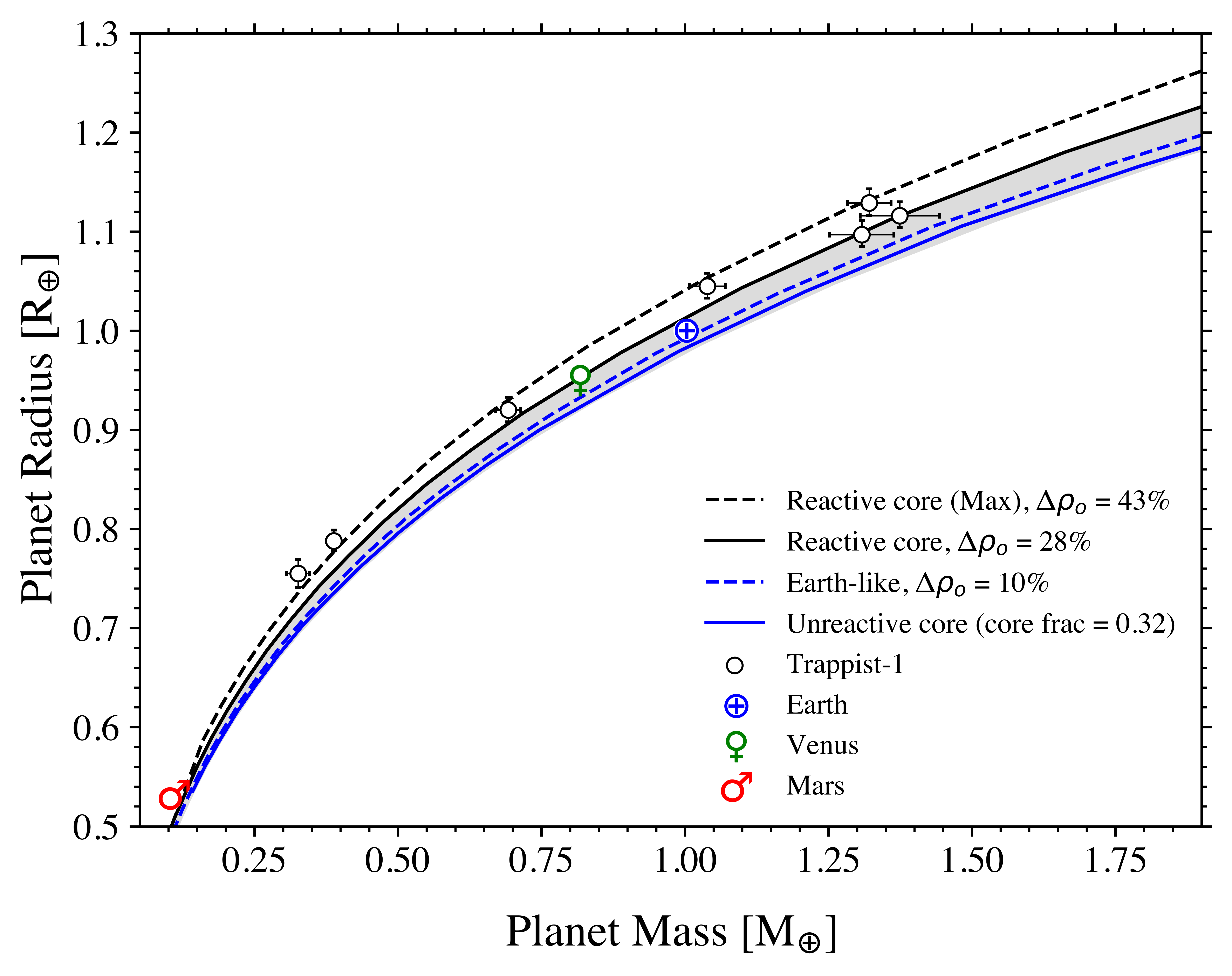}
\caption{Closeup of Figure \ref{Fig:M_vs_R} showing the details of the relationships between Solar System rocky planets, those of Trappist-1 b, c, d, e, f, g, and h after \cite{Agol2021}, compared with mass-radius relations for various density deficits due to H and O in metal cores. Curves are the same as those shown in Figure \ref{Fig:M_vs_R}. Inner Solar-system planets are shown for comparison. Interestingly, the various under-dense Trappist-1 planets can be explained by standard core fractions when accounting for light elements in the core found in our reactive core models. Our results therefore, may offer an alternative explanation for the lower than expected densities of the planets in the Trappist-1 system \citep[e.g.][]{Agol2021}.}  
\label{Fig:M_vs_R_closeup}
\end{figure}

Also shown in Figures \ref{Fig:M_vs_R} and \ref{Fig:M_vs_R_closeup}, as the shaded region, is the $1\sigma$ range of super-Earth  bulk densities derived from planetary evolution and atmospheric mass-loss models that successfully reproduce the radius valley in the exoplanet size distribution \citep{rogersj2020a}. Comparing the data and shaded region with our reactive and unreactive mass-radius relations in Figure \ref{Fig:M_vs_R} shows that both reactive and unreactive core solutions are fully consistent with current observations of super-Earth masses and radii.

In addition, our curves showing the effects of density deficits due to H and O mixed with Fe in cores illustrate that any true variability in densities among super-Earths, if it exists, could be due to variability in the light elements in the cores, rather than large water fractions as has been suggested in past works \citep[e.g.][]{UDH2018}.

For example, the two inner planets around the K-dwarf star HD 219134 have densities indicative of rock and, possibly, metal interiors (Figure \ref{Fig:M_vs_R}). The differences between the two planets, b and c, has been variously attributed to differences in the mass fraction of dense metal cores \citep{Gillon2017}, or differences in rock composition resulting from incomplete condensation in the protoplanetary disk \citep{Dorn2019}. Figure \ref{Fig:M_vs_R} shows that the range in densities can also be explained by differences in the composition of the metal cores, with planet HD 219134 b, the more massive of the two, potentially having more H and O in its core compared with the less massive HD 219134 c. Assuming Earth-like core mass fractions, the density of the former is explained by a density deficit in the Fe core of $\sim 30\%$, corresponding to 2 wt.\% H and 10 wt.\% O in the core, while that of the latter can be explained by an Earth-like core composition, with a density deficit relative to Fe of about $10 \%$, corresponding to $0.7$ wt. \% H and $3.4$ wt. \% O in the core. If the bulk density differences of HD 219134 b and c are indeed due to different amounts of light elements in their cores then this would be evidence that they likely formed with primordial hydrogen envelopes, which were subsequently lost by either core-powered mass-loss \citep{gupta2019a} or photo-evaporation \citep{rogersj2020a}.

The rocky planets around Trappist-1 exhibit lower densities than Earth (Figure \ref{Fig:M_vs_R_closeup}).  The lower densities have been explained as being the result of lower mass fractions of metal cores, no cores, or perhaps water-rich surface layers \citep[e.g.,][]{Agol2021}.  Figure \ref{Fig:M_vs_R_closeup} shows that these lower densities could also be explained with Earth-like core fractions if the cores have $~ 20$\% to $40\%$ density deficits due to H and O in their metal. Such concentrations of H and O are tenable since these planets could have accrued transient primary atmospheres with masses at per cent levels  \citep{ginzburg2016a}.  \begin{ed} Indeed, our proto-Earth calculations provide reasonable analogues.  For a $1\%$ by mass primary atmosphere of H$_2$, we obtain a $17\%$ density deficit for the core for a silicate-metal equilibration temperature of 3000 K and a $\sim 40\%$ density deficit for an equilibration temperature of $6000$ K.  These values bracket those required to explain the densities of the Trappist-1 planets.\end{ed}     

The effects of hydrogen and other light elements in the cores of differentiated planets adds another dimension that may potentially confound attempts to constrain the mass fractions of condensed water or secondary atmospheres for under-dense super-Earths, adding yet another degeneracy to an already under-determined problem \citep[e.g][]{SeRo2010,Crida2018, dorn2017b}.

\subsection{The Mass-Radius Relations of Sub-Neptunes and Observational Tests}

\begin{figure*}
\centerline
{\includegraphics[width=0.8\textwidth]{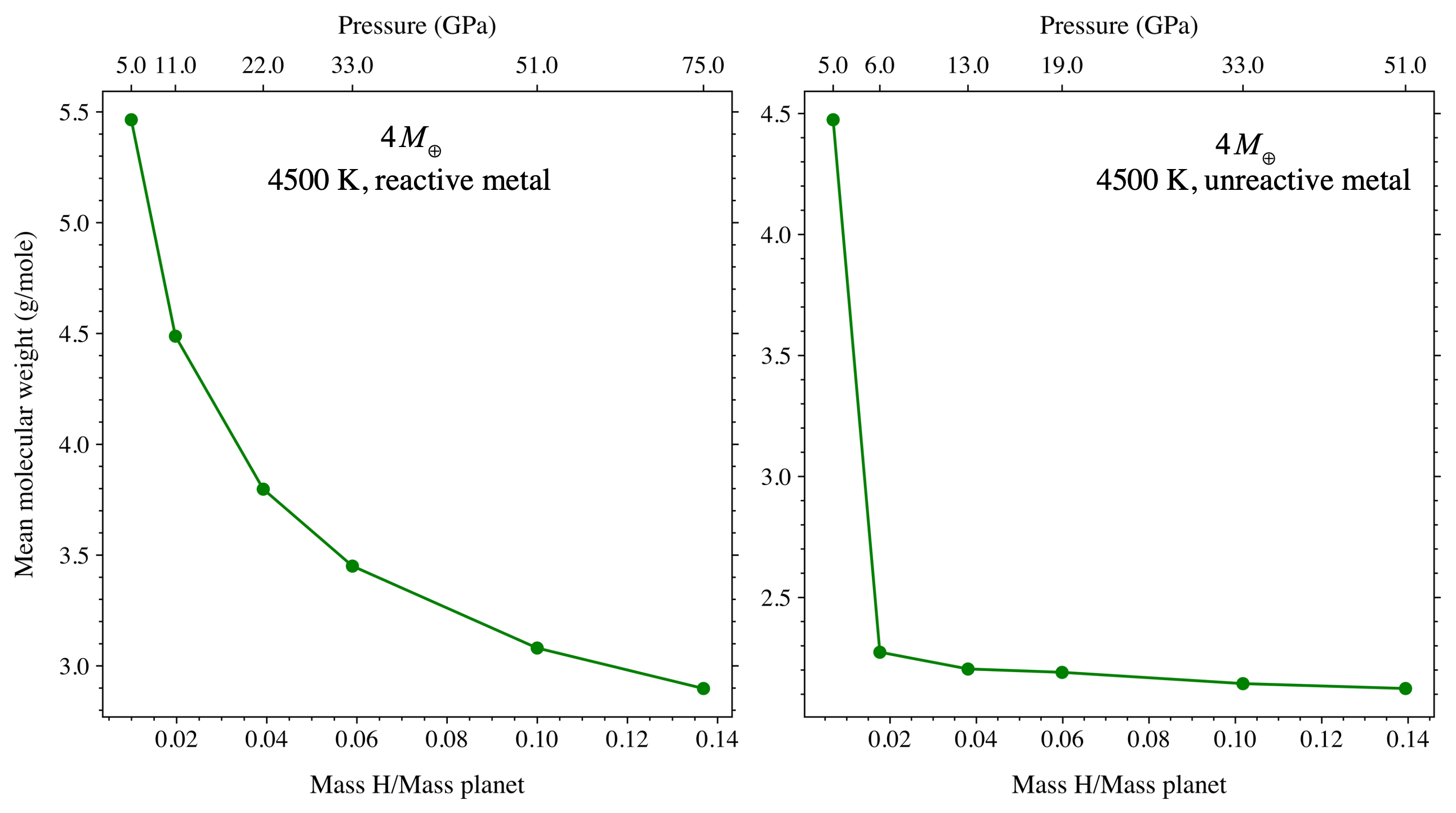}}\hfil
\caption{Mean molecular weights for atmospheres based on equilibration of a  4$M_{\Earth}$ planet at 1400 K as a function of  total H ranging from 1\% to 14\% by weight.  On the left is the variation with total mass fraction for the reactive metal case at 4500 K, and on the right is the same for the unreactive metal case.}
\label{Fig:molecular_weights}
\end{figure*}

Allowing for equilibrium chemistry between the atmosphere, mantle and core has also several interesting implications for sub-Neptunes. Similar to the super-Earths, including equilibrium chemistry will modify the mass-radius relation for sub-Neptunes. Specifically, it can lead to a modification of core densities, mean-molecular weight of the atmosphere and atmospheric structure. In addition, including equilibrium chemistry likely also affects their thermal evolution and mass-loss histories. We discuss these in turn below.

Perhaps the most significant effect of equilibrium chemistry on sub-Neptune mass-radius relations is the increases in molecular weights of their atmospheres (Figure \ref{Fig:molecular_weights}). We find that the mean molecular weight averaged over the whole atmosphere typically is increased by about a factor of more than two, from 2 amu, for a pure $\rm{H_2}$ atmosphere, to $>$ 5 amu.  In all cases, the higher mean molecular weights of the atmospheres compared to that of pure H$_2$ are attributable largely to the presence of SiO, which is a result of evaporation of the magma ocean, and, in the case of the unreactive metal models, H$_2$O (see Figures \ref{Fig:atmosphere_reactive} and \ref{Fig:atmosphere_unreactive}).

\begin{ed}
Recent work investigating the consequences of SiO vapour in sub-Neptune atmospheres showed that silicate vapor in a hydrogen-dominated atmosphere acts as a condensable species \citep{Misener2022}. Since the silicate vapor deceases in abundance with altitude a mean molecular weight gradient develops, which inhibits convection at temperatures above $\sim 4000$~K, inducing a near-surface layer where, depending on the exact conditions, heat transport is either dominated by radiation or convection. This none-convective layer decreases the planet's total radius compared to a planet with the same base temperature and a convective, pure H/He atmosphere. The presence of silicate vapor, therefor, can have a significant effects on the inferred envelope masses inferred from mass-radius measurements and for the structure and thermal evolution of envelopes of sub-Neptunes \citep{Misener2022}. 
\end{ed}

\begin{ed}
Exciting is the prospect, that by probing the atmospheric composition of sub-Neptunes, one can potentially distinguish between the reactive and unreactive metal cases. The opportunity for such observational tests arises because SiO is the most abundant species by mole faction after hydrogen in the atmosphere for planets where metal can fully participate in the equilibrium chemistry, whereas water is the most abundant species by mole faction after hydrogen in the atmosphere for planets with unreactive metal (see Section \ref{chemat} for details).
\end{ed}


\subsection{Implications for the Formation and Evolution of super-Earths and sub-Neptunes}

\begin{ed}
The result that large fractions of H$_2$ are absorbed as melt by the magma ocean in the unreactive core case for a total hydrogen mass fraction of 1\% (Figure \ref{Fig:Hfraction_unreactive}) agrees with previous work by \cite{Chachan2018} who investigated a range of planet masses but fixed their hydrogen mass fractions at about 1\%.  These authors only considered a single chemical reaction describing hydrogen solubility in the magma and therefore had no mechanisms to produce or consume water.   

\end{ed}

Unlike \cite{Kite2019}, who used ostensibly similar assumptions regarding hydrogen solubility in the magma ocean \citep{Hirschmann2016} but different means of extrapolations to higher pressures (Figure \ref{Fig:R15_solubility}), we find no evidence that large fractions of the total hydrogen budget can be dissolved in the magma for hydrogen mass fraction larger than $\sim$ 1\% (i.e. for pressures greater than $\sim 5$ GPa, Figure \ref{Fig:Hfraction}). The relatively low abundance of planets with sizes between Neptune and Saturn is therefore probably not due to large amounts of hydrogen being absorbed by the mantle, but instead due to the on-set of runaway gas accretion which rapidly turns a sub-Neptune into a gas giant \citep[e.g.][]{Pollack1996}. In addition, as the envelope accretion of sub-Neptunes is not supply/hydrogen-limited, but limited by the envelope's ability to cool and contract such that it can make room for more hydrogen to enter the Bondi sphere \citep{lee2015a,ginzburg2016a}, any hydrogen dissolved in the magma would be rapidly re-supplied from the disk, thereby not leaving an observational signature in the sub-Neptune radii.

Finally, we would like to stress that because atmospheric mass-loss, thermal evolution, and equilibrium chemistry for a planet evolve together over time, self-consistent modelling is needed, especially because of the temperature and pressure dependencies of the chemistry, to make any concluding inference about a planet's final detailed interior and atmospheric composition.

\section{Summary \& Conclusions}
In this work, we investigated how the coexistence of hydrogen-rich atmospheres,  magma oceans, and metal cores affect the physical and chemical evolution of super-Earths and sub-Neptunes. We used a basis set of reactions depicting chemical equilibrium between the atmospheres, silicate mantles, and  iron-rich metal cores. To account for the possibility of differentiation and limited bulk transport in a given planet, we investigated both a ``reactive metal" case, in which the metal, not necessarily coalesced into a core, participates fully in the equilibrium chemistry, and an ``unreactive metal" case, where the metal core is chemically isolated from the rest of the planet. The equilibrium chemistry in the ``reactive metal" case does not distinguish between a differentiated planet with a metal core or an undifferentiated planet where metal is co-existing with the silicates in the mantle. Key findings of this work are as follows:

\begin{itemize}

\begin{ed}
  \item When Fe metal can fully participate in equilibrium chemistry, hydrogen and oxygen combined make up a few to tens of per cent of the core/iron metal.
  
  \item The presence of light elements, especially H and O, in the metal results in a density deficit in the metal core, thereby modifying the mass-radius relation of rocky planets, adding yet another degeneracy to an already under-determined problem of inferring composition from planet mass and radius measurements.
  
  \item Molar concentrations in the envelopes of planets with reactive metal are  H$_2$ $>$  SiO $>$ CO $\sim$ Na $\sim$ Mg $>$ H$_2$O $>>$ CO$_2$ $\sim$ CH$_4$ $>>$ O$_2$ while for the unreactive metal case H$_2$O becomes the second most abundant species on a molar basis, after H$_2$.  This may provide an arbiter for the two scenarios amenable to observation. 
  
  \item The water abundance in the atmosphere exceeds that in the mantle by at least an order of magnitude in both the reactive and unreactive metal scenarios.
  
  \item  The water content in the silicate mantle is about 0.01 wt\% and 0.1 wt\% in the reactive and unreactive metal core cases, respectively, limiting the H$_2$O that might be outgassed from the magma ocean in a future super-Earth. Less dissolved water in the reactive core case is due to sequestration of both H and O in the Fe-rich metal.
  
  \item  The total hydrogen budget of most sub-Neptunes can be, to first order, well estimated from their atmospheres alone, as the atmospheres typically contain more than 90\% of all H except where H is $<$ 2\% of the mass of the planet or equilibration temperatures are $>$ 5500 K.
  
  \end{ed}

  
\end{itemize}

It is important to stress here that the amounts of light elements in the mantle and core determined from equilibrium chemistry alone should be regarded as upper limits, as their actual abundance will likely be limited by their bulk transport though the planet.
In addition, the light element abundances in the core and mantle are also limited by the requirement that the metal core and silicate mantle densities need to be greater than, or equal to, that of the mantle at the core-mantle boundary and that of the atmosphere at the atmosphere-mantle boundary, respectively. We showed above that this imposes a maximum density  deficit  for  the  core  of about  43\%, corresponding to 5\% by mass H in the metal or, more realistically, a combination of 3\% H and 14\% O by mass.

The work presented here is meant as a first step towards understanding the compositional, chemical, structural, and thermal evolution of planets born with initial hydrogen dominated envelopes comprising a few to several percent of their total mass. We made many simplifying assumptions in this paper to achieve this goal, several of which should be relaxed in future work.


\begin{ed}
Future models should include estimates of the nature of H and O-rich Fe-bearing metals at extreme pressures, and the nature of energy and mass transfer between the surface of magma oceans and the core-mantle boundaries. \end{ed}


Future work should model the full evolution of a given planet over time by coupling our chemical equilibrium model with atmospheric mass-loss calculations and by evolving them together over time. This work will be needed in order to determine the composition and chemistry of super-Earths, and, as an extension, their suitability for habitability, as they appear to be born from sub-Neptunes that are stripped of their primordial atmospheres over time.

\section*{Appendix: Thermodynamic data}

Here we describe the sources of thermodynamic data used to calculate standard-state molar Gibbs free energies of reaction, $\Delta \hat G_{{\rm{rxn}}}^ \circ$.  Where possible we sought to recover internally consistent free energies of formation of the various species $i$ in the molten silicate and $j$ in the molten metal ($\Delta \hat G_i^{\rm o,silicate}$ and $\Delta \hat G_j^{\rm o,metal}$, respectively, where $^{\rm o}$ indicates the pure species at 1 bar and temperature of interest). Ideal mixing is used for all phases.  We compared our results where we use 1-bar standard state free energies of formation for the melt species to those where corrections for pressure are made for these species. We found only relatively minor effects on free energies of reaction when correcting melt species for pressure, comparable in magnitude to the uncertainties in the thermodynamic data. In addition, partial molar volumes are not available for all of the light elements in the metal.  For these reasons, no corrections for pressure were made for the melt species.  

\medskip
\noindent R1: MAGMA code,  \cite{Fegley1987}

\medskip
\noindent R2: $\Delta \hat G_{\rm rxn,R2}^{\rm o}= - 2.303 RT (0.364-16520/T)$ after \cite{Badro2015}.

\medskip
\noindent R3: NIST

\medskip
\noindent R4: $\Delta \hat G_{{\rm{rxn,R4}}}^ \circ=-(\Delta \hat G_{\rm rxn,Ox}^{\rm o}+\Delta \hat G_{\rm rxn,R2}^{\rm o})$ where ``Ox" refers to the reaction FeO = Fe + O (O in metal) from \cite{Badro2015}.  After  correction for a sign typographical error for the $\Delta \hat H_{\rm rxn}^{\rm o}/R$ term (Julien Siebert, pers. comm.),  $\Delta \hat G_{\rm rxn,Ox}^{\rm o}= - 2.303 RT (2.736-11439/T)$. 

\medskip
\noindent R5:  $\Delta \hat G_{\rm rxn, R5}^{\rm o}= \Delta \hat G_{\rm H_2}^{\rm o,silicate}-2\Delta \hat G_{\rm H}^{\rm o,metal}$. $\Delta \hat G_{\rm H_2}^{\rm o,silicate}$ for peridotite melt at $T$ and 1 bar was obtained from the free energy of the reaction H$_{2, { \rm gas}}$ = H$_{2, {\rm melt}}$  after  \cite{Hirschmann2012}, and $\Delta \hat G_{\rm H_2}^{\rm o}$ for gas from NIST.  The $\Delta \hat G_{\rm H}^{\rm o}$ for metal was obtained from  the reaction Fe + H$_2$O$_{\rm melt}$ = FeO + 2H  by regression of $\ln(K_{\rm eq})$ vs. $1/T$ reported by \cite{Okuchi1997}, yielding $\Delta \hat G_{\rm rxn, Okuchi97}^{\rm o} = 143589.7-69.1 T$ (J/mole),  $\Delta \hat G_{\rm H_2O}^{\rm o}$ for silicate melt from the solubility data of \cite{Moore1998} for the reaction H$_2$O$_{\rm gas}$ = H$_2$O$_{\rm melt}$ (see discussion for R16), and the standard-state  free energies of formation for H$_2$O gas, liquid FeO, and Fe from NIST. From these values, $\Delta \hat G_{\rm H}^{\rm o,metal} = 1/2( \Delta \hat G_{\rm rxn, Okuchi97}^{\rm o}- \Delta \hat G_{\rm FeO}^{\rm o} +\Delta \hat G_{\rm Fe}^{\rm o}+\Delta \hat G_{\rm H_2O}^{\rm o,silicate})$.

\medskip
\noindent R6: $\Delta \hat G_{\rm rxn, R6}^{\rm o}$ is taken from the Magma code for the reaction  Fe$_2$SiO$_4$ = 2FeO + SiO$_2$,  \cite{Fegley1987}.  We avoided using published estimates for the free energy of formation for molten FeSiO$_3$ because when paired with our data for other melt species they produce erratic solutions. The precise value for the free energy of reaction for R6 is not critical to our results.

\medskip
\noindent R7:  Calculated from the Gibbs free energy values referenced above and $\Delta \hat G_{\rm Si}^{\rm o}$.  The latter is obtained by difference from the $\Delta \hat G_{{\rm{rxn}}}^ \circ$ for the reaction $1/2$SiO$_2$ + Fe$_{\rm metal}$ = FeO + $1/2$Si$_{\rm metal}$ (R2) from  \cite{Badro2015}, making R2 and R7 self consistent. 

\medskip
\noindent R8: NIST

\medskip
\noindent R9: NIST

\medskip
\noindent R10: NIST

\medskip
\noindent R11: NIST

\medskip
\noindent R12: NIST

\medskip
\noindent R13: NIST

\medskip
\noindent R14: NIST

\medskip
\noindent \begin{ed}
R15:  $\Delta \hat G_{\rm rxn, R15}^{\rm o}= \Delta \hat G_{\rm H_2}^{\rm o, silicate}  - \Delta \hat G_{\rm H_2}^{\rm o, gas}$, where $\Delta \hat G_{\rm H_2}^{\rm o, silicate}$ at $T$ and 1 bar was obtained from the free energy of the reaction H$_{2, { \rm gas}}$ = H$_{2, {\rm melt}}$  for peridotite from \cite{Hirschmann2012}.  We used a constant melt/vapor distribution coefficient, $K_{\rm D}=x_\mathrm{H_2, silicate}/x_\mathrm{H_2, gas}$.  The resulting dependence of the equilibrium constant, $K_{\rm eq}$, on the inverse of pressure, where $K_{\rm eq}=x_\mathrm{H_2, silicate}/(x_\mathrm{H_2, gas}P)$$=k_{\rm D}/P$, is consistent with the experimental data and avoids unconstrained extrapolation of the negative dependence of $\ln(K_{\rm eq})$ on $P$ obtained from low-pressure experiments  \citep{Hirschmann2012}. Accordingly, the curves for reference in Figure (\ref{Fig:R15_solubility}) are obtained using $K_{\rm D}= -\Delta \hat G_{\rm rxn, R15, 3{\rm GPa}}^{\rm o}/(RT)-\ln(P/{\rm 3GPa})$.
\end{ed}

\smallskip
\noindent R16:  $\Delta \hat G_{\rm rxn,R16}^{\rm o}=\Delta \hat G_{\rm H_2O}^{\rm o,silicate}-\Delta \hat G_{\rm H_2O}^{\rm o, gas}$.
$\Delta \hat G_{\rm H_2O}^{\rm o,silicate}$ was obtained from the solubility data of \cite{Moore1998}  for the reaction H$_2$O$_{\rm gas}$ = H$_2$O$_{\rm melt}$, consistent with the data for R5 and R7.  The data of
Moore et al.\  are fit well by $x_{{\rm H}_2{\rm O}}^{\rm silicate}=\exp(1/2(-\Delta \hat H_{\rm rxn}^{\rm o}/(RT) + \Delta \hat S_{\rm rxn}^{\rm o}/R + 1.17 \ln(P/{\rm bar})))$, where $\Delta \hat H_{\rm rxn}^{\rm o}/R=-2565$ and $\Delta \hat S_{\rm rxn}^{\rm o}/R = -14.21$.  

\medskip
\noindent R17: Assumed to be $1/3$ the solubility of CO$_2$ based on \cite{Hirschmann2016}

\medskip
\noindent R18: \cite{Pan1991}

\section*{Acknowledgements}
This research has made use of the NASA Exoplanet Archive, which is operated by the California Institute of Technology, under contract with the National Aeronautics and Space Administration under the Exoplanet Exploration Program. HES thanks William Misener and Akash Gupta for insightful discussions and gratefully acknowledges support from the National Aeronautics and Space Administration under grant No. 80NSSC18K0828. EDY acknowledges support from NASA Emerging Worlds grant No. 80NSSX19K0511.

\bibliographystyle{aasjournal}


\end{document}